\documentclass[aps,amssymb,amsmath,notitlepage,superscriptaddress,pra,twocolumn,nofootinbib]{revtex4-1}

\usepackage{setspace}
\usepackage{amsfonts}
\usepackage{graphicx,graphics,epsfig,times,bm,bbm,amssymb,amsmath,amsfonts,mathrsfs}
\usepackage[normalem]{ulem}
\usepackage{wrapfig}
\usepackage{setspace}
\usepackage{subfigure}
\usepackage{dsfont}
\usepackage{braket}
\usepackage[pdfstartview=FitH]{hyperref}
\usepackage{upgreek }
\usepackage{tikz}
\usepackage{natbib}
\usepackage{chngcntr}
\usepackage{stmaryrd}
\usepackage[makeroom]{cancel}

\newtheorem{theorem}{Theorem}

\newcommand{\bes} {\begin{subequations}}
\newcommand{\ees} {\end{subequations}}
\newcommand{\bea} {\begin{eqnarray}}
\newcommand{\eea} {\end{eqnarray}}
\newcommand{\beq}{\begin{equation}}
\newcommand{\eeq}{\end{equation}}

\def\>{\rangle}
\def\<{\langle}
\def\Tr{\textrm{Tr}}

\def\B{\textrm{R}}
\def\SB{\textrm{SR}}

\renewcommand{\min}{\textrm{min}}
\renewcommand{\max}{\textrm{max}}

\newcommand{\ignore}[1]{}

\def\Tr{\textrm{Tr}}

\def\E{\textrm{eq}}

\def\I{\textrm{I}}
\def\S{\textrm{S}}
\def\B{\textrm{B}}

\def\>{\rangle}
\def\<{\langle}
\def\Tr{\textrm{Tr}}

\def\B{\textrm{E}}
\def\SB{\textrm{SE}}

\def\LS{\textrm{LS}}

\renewcommand{\min}{\textrm{min}}
\renewcommand{\max}{\textrm{max}}

\begin{document}

\title{Exponential suppression of decoherence and relaxation of quantum systems using energy penalty}

\author{Iman Marvian}
\affiliation{Research Laboratory of Electronics, Massachusetts Institute of Technology, Cambridge, MA 02139}

\begin{abstract}

One of the main methods for protecting quantum information against decoherence  is to encode information in the ground subspace (or the low energy sector)  of a Hamiltonian with a large energy gap which penalizes errors from environment. The protecting Hamiltonian is chosen such that its degenerate ground subspace is an error detecting code for the errors caused by the interaction with the environment.   We consider environments with arbitrary number of local sites, e.g. spins, whose interactions among themselves are local and bounded. Then, assuming the system is interacting with a finite number of sites in the environment, we  prove that,  up to second order with respect to the coupling constant, decoherence and relaxation are suppressed by  a factor which grows exponentially fast with the ratio of energy penalty to the norm of local interactions in the environment. The state may, however,  still evolve unitarily inside the code subspace due to the Lamb shift effect.  
In the context of adiabatic quantum computation,  this means that the evolution inside the code subspace is effectively governed by a renormalized Hamiltonian. 
The result is derived from first principles, without use of master equations or their assumptions, and holds even in the infinite temperature limit.  We also prove that   unbounded or non-local  interactions in the environment at sites  far from the system do not considerably modify the exponential suppression.
\end{abstract}
\maketitle

\textit{Introduction}.---
Protecting quantum information against errors and decoherence  is a major challenge for the progress of quantum information technology.  In the last couple of decades, several methods have been proposed to overcome this challenge (See e.g. \cite{shor1995scheme, Gottesman:97b, Zanardi:97c, LidarDFS,  Viola:99,Viola:98, PhysRevLett.100.160506, kitaev2003fault, HQC, PhysRevA.74.052322}). A ubiquitous approach for passive suppression of errors, which can be used in conjunction with other methods, is to store information in the ground subspace of a degenerate Hamiltonian with a large energy gap $E_\text{gap}$ which penalizes errors from the environment \cite{kitaev2003fault, HQC, PhysRevA.74.052322}. This approach is particularly important in the context of Adiabatic Quantum Computation \cite{FarhiAQC:00,PhysRevA.74.052322}, where the  resources required to  implement the standard error correction  algorithms, such as fresh ancillas and measurements, are not available.   The protecting Hamiltonian is chosen such that its ground subspace, or the \emph{code subspace}, is an \emph{error detecting code} for the errors caused by the interaction with the environment \cite{ Lidar-Brun:book, knill2000theory}. This condition guarantees  that states in the code subspace are effectively decoupled from  the environment in the limit of large energy penalty \cite{bookatz2015error, marvian2015quantum, marvian2014quantum}.

It is not, however, clear that in practice, where $E_\text{gap}$ is finite,  to what extent and under what circumstances this method could be useful. This question has been recently studied by  Bookatz et al. in \cite{bookatz2015error}, where they show that, if the system-environment coupling is norm-bounded, local and weak, and the interactions inside the environment are also norm-bounded and local, then up to the second order with respect to the coupling constant, the fidelity loss is upper bounded by $E^{-2}_\text{gap} $ times a quadratic function of time. A similar bound has also been obtained in \cite{marvian2015quantum}, using a  non-perturbative exact approach.  Bookatz et al. \cite{bookatz2015error} also performed long-time  numerical simulations for a small environment and observed that, although as predicted by the above bound, the speed of fidelity loss is suppresses by $E^{-1}_\text{gap}$,  the state remains inside the code subspace for much longer times.  

In this Letter we study error suppression with finite energy penalty for the case of environments formed from arbitrary number of local sites, e.g. spins, interacting via local and bounded interactions with each other (i.e. the same assumptions made by Bookatz et al. \cite{bookatz2015error}). An important class of examples of this type of environments are spin-bath models \cite{Prokofev:00}. Similar to   \cite{bookatz2015error} and \cite{Alicki:02}, we  use a perturbative approach to study the fidelity loss. Then, starting from first principles, we rigorously prove that, up to the second order with respect to the coupling constant, decoherence is  suppressed by a factor which grows exponentially fast  with the ratio of energy penalty $E_\text{gap}$ to the norm of local interactions in the environment.  However, our analysis reveals that, even though decoherence is slowed down exponentially,  the state still evolves unitarily inside the code subspace due to the Lamb shift effect. Therefore, to retrieve the initial state one needs to correct the effect of this unitary evolution. Ignoring this unitary evolution in the code subspace, which has not been noticed before, results in a much weaker suppression of fidelity loss. That is,  instead of exponential suppression, the fidelity loss will be suppressed only by $E^{-1}_\text{gap}$, as observed in  \cite{bookatz2015error} and \cite{marvian2015quantum}.  
Our result also explains the numerical observation of \cite{bookatz2015error} regarding the strong suppression of leakage outside  the code subspace. 


To prove this result on exponential suppression of errors, we introduce two other new results, which are of independent interest. First, we present a general perturbative theory of error suppression, and prove a new theorem, which  establishes a framework for understanding how error suppression with energy penalty works in the perturbative regime. Second,  we find a bound on the decay of the power spectral density of local observables at high frequencies. This bound  formalizes the intuition that  the high-frequency oscillations of  many-body systems have negligible effects on local observables.

\textit{Error suppression with energy penalty}.--- Consider a system S with Hamiltonian $H_\S$, and let the \emph{code subspace} $\mathcal{C}$ be the ground subspace of $H_\S$, which is separated from the rest of its spectrum by  gap $E_\text{gap}>0$. Without loss of generality, we assume the ground state energy is zero. Let $\Pi_\mathcal{C}$ be the projector onto $\mathcal{C}$.  Suppose the system S is initialized  in state $\rho_\S$ in $\mathcal{C}$, and  then at $t=0$  is coupled to an environment E (bath) with Hamiltonian $H_\B$ via a coupling Hamiltonian $\lambda H_\I$, where the dimensionless coupling constant $\lambda$ determines the coupling strength. We assume the environment is initially uncorrelated with the system S, and is in the equilibrium state $\rho_\B$ (i.e. $[\rho_\B,H_\B]=0$)  which is not necessarily a thermal state. Therefore, the joint initial state of system and environment at $t=0$ is $\rho_\S\otimes\rho_\B$, and the  total Hamiltonian at $t>0$ is  $H_\text{S}+\lambda {H}_\text{I}+H_\B\ $.

Consider the decomposition of the coupling Hamiltonian as $\lambda H_\I=\lambda\sum_i S_i\otimes B_i$ with linearly independent $\{S_i\}$ and $\{B_i\}$. In the theory of quantum error correction \cite{Lidar-Brun:book} the subspace $\mathcal{C}$ is called an \emph{error detecting code} for the set of errors $\{S_i\}$ if  $\Pi_\mathcal{C} S_i \Pi_\mathcal{C}\propto \Pi_\mathcal{C}$ for all $S_i$.  This condition can be interpreted as the quantum version of the classical error detection condition, which guarantees that the errors do not mix different codewords. Interestingly, in the context of quantum error suppression, this condition finds a different interpretation. Note that for $\lambda H_\I=\lambda\sum_i S_i\otimes B_i$ with linearly independent $\{S_i\}$ and $\{B_i\}$ this condition is equivalent to
\beq\label{cond}
(\Pi_\mathcal{C}\otimes I_\B) H_\I (\Pi_\mathcal{C}\otimes I_\B) =\Pi_\mathcal{C}\otimes O_\B\ ,
\eeq
where $I_\B$ and $O_\B$, are respectively the identity operator and an arbitrary operator  on the environment Hilbert space. Then, in the limit  $E_\text{gap}\rightarrow\infty$, the left-hand side of Eq.(\ref{cond})  is the effective interaction between states in the code subspace and the environment in the first order degenerate perturbation theory.  Therefore, if this equation holds then  states in $\mathcal{C}$ remain unaffected by the environment in the limit  $E_\text{gap}\rightarrow\infty$. In the following we are interested to evaluate the effectiveness of this method in a more realistic setting, where $E_\text{gap}$ is finite and the  coupling is weak,  $\lambda\ll 1$. 

\textit{Perturbative theory of error suppression}.---
To focus on the main ideas and simplify the presentation we assume the interaction  is in the form $\lambda H_\I=\lambda S\otimes B$. Extension  to the case of general interaction  $H_\I=\lambda \sum_i S_i\otimes B_i $ is straightforward. Let $H_\S=\sum_n E_n\ \Pi_{E_n}$ be the spectral decomposition of $H_\S$,  and $\Omega=\{E_n-E_m\}$ be the corresponding set of frequencies (Throughout this paper we assume $\hbar=1$). For any frequency $\mu\in\Omega$, let $S_\mu=\sum_n \Pi_{E_n} S \Pi_{E_n+\mu}$ be  the component of the system operator $S$ in frequency $\mu$. Finally, let  $\rho_\S(t)$  and $\tilde{\rho}_\S(t)=U_\S^\dag(t) \rho_\S(t) U_\S(t)$, be  the reduced state of system S at time $t$ in the lab frame and in the interaction picture respectively, where $U_\S(t)=e^{-i H_\S t}$. Then, as it is shown in the supplementary material, following  \cite{PhysRevA.40.4077,Alicki:02,Majenz:2013qw}, by  truncating the Dyson series   in the interaction picture we find
\begin{align}\label{Eqmotion}
\tilde{\rho}_\S(t)&=\rho_\S-i [\lambda F_\I(t)+ \lambda^2 F_\LS(t) ,\rho_\S] + \lambda^2 \Phi_t(\rho_\S)+\mathcal{O}(\lambda^3) \ . 
\end{align}
Here $i [ \lambda F_\I(t)+\lambda^2 F_\LS(t) ,\rho_\S]$ describes a Hamiltonian evolution  due to the interaction with the environment. In particular, the first order term $\lambda F_\I(t)=\lambda \int^t_0 ds\ U_\S^\dag(s)  \Tr_\B(H_\I \rho_\B) U_\S(s)$ is basically the effect of the average Hamiltonian $\lambda \Tr_\B(H_\I \rho_\B)$, which is sometimes  absorbed in $H_\S$.  The second order term $\lambda^2 F_\LS(t)$, known as \emph{Lamb Shift} effect, is determined by the autocorrelation function of the environment operator $B$ (See Supplementary Material). On the other hand, the  superoperator  $\Phi_t$ describes the part of evolution which could be dissipative, and is given by
\begin{align}\label{expansion}
\Phi_t(\rho)\equiv \sum_{\mu,\mu'\in\Omega} b_{\mu\mu'}(t) \big[ S^\dag_{\mu'} \rho S_\mu-\frac{1}{2} \{S_\mu S^\dag_{\mu'}, \rho\}  \big]\ ,   
\end{align}
where $b_{\mu\mu'}(t)=\int_0^t\int_0^t ds_1 ds_2  e^{i(\mu' s_2-\mu s_1)} C_B(s_1-s_2)$,  and $C_B(t)=\Tr(\rho_\B e^{i H_\B t}B e^{-i H_\B t} B)$ is the autocorrelation function of $B$.

  
We are interested in the Uhlmann \emph{fidelity} \cite{nielsen2000quantum, Uhlmann, Fidelity_Jozsa} of state $\rho_\S(t)$ with the initial state $\rho_\S$. Recall that the fidelity of two positive operators $\sigma_1$ and $\sigma_2$ is given by $\text{F}(\sigma_1,\sigma_2)\equiv \Tr(\sqrt{\sqrt{\sigma_1} \sigma_2 \sqrt{\sigma_1}})$. 
 Since the initial state $\rho_\S$ is  in the code subspace, its fidelity with any arbitrary state $\sigma$ only depends on  $\Pi_\mathcal{C}\sigma \Pi_\mathcal{C}$,  that is the restriction of  $\sigma$ to $\mathcal{C}$. More precisely, 
$\text{F}(\rho_\S, \sigma)=\text{F}(\rho_\S,\Pi_\mathcal{C} \sigma \Pi_\mathcal{C})$. Therefore, to find the fidelity of state ${\rho}_\S(t) $ with the initial state $\rho_\S$ we can focus on the dynamics of $\Pi_\mathcal{C} {\rho}_\S(t) \Pi_\mathcal{C}=\Pi_\mathcal{C}\tilde{\rho}_\S(t) \Pi_\mathcal{C}$. This simple observation is useful in the following, and in particular  implies that we can neglect  all the terms with $\mu\neq \mu'$ in the expansion of $\Phi_t$ in Eq.(\ref{expansion}), because they  vanish in $\Pi_\mathcal{C} \Phi_t(\rho_\S) \Pi_\mathcal{C} $ (even though we have not made the rotating wave approximation).

Using this observation and in the light of Eq.(\ref{Eqmotion}) we can clearly see the importance of the error detection condition  Eq.(\ref{cond}): First, it implies that the effect of $\lambda \Tr_\B(H_\I \rho_\B)$   vanishes inside $\mathcal{C}$, and hence $\Pi_\mathcal{C} {\rho}_\S(t) \Pi_\mathcal{C}$ does not have any first order term in $\lambda$. Second, it implies that in the expansion of $ \Pi_\mathcal{C}\Phi_t(\rho_\S) \Pi_\mathcal{C}$ obtained from Eq.(\ref{expansion}) the terms with zero frequency $\mu,\mu'=0$  cancel each other  (This follows from the fact that both $S_0$ and $\rho_\S$ commute with $\Pi_\mathcal{C}$, and $\Pi_\mathcal{C}S_0 \Pi_\mathcal{C}=\Pi_\mathcal{C}S \Pi_\mathcal{C}  \propto \Pi_\mathcal{C} $). 
Next, using the fact that the system is initially in the ground subspace, we find that the only frequencies $\mu\in\Omega$ which contribute in   $\Pi_\mathcal{C}\Phi_t(\rho_\S) \Pi_\mathcal{C}$ are frequencies $\mu \ge E_\text{gap}$. To summarize, the error detection condition implies that up to  $\mathcal{O}(\lambda^2)$, 
\begin{align}\label{eqasd}
\Pi_\mathcal{C}{\rho}_\S(t)\Pi_\mathcal{C}&=\rho_\S-i \lambda^2 [ \Pi_\mathcal{C}F_\LS(t)\Pi_\mathcal{C} ,\rho_\S]\ -\frac{\lambda^2}{2} \{ \rho_\S, A(t)\}\ ,
\end{align}
where $A(t)= \sum_{\mu\ge E_\text{gap}} b_{\mu,\mu}(t) \ \Pi_\mathcal{C}  S_{\mu}S^\dag_\mu \Pi_\mathcal{C}$.  But the term $-i \lambda^2 [ \Pi_\mathcal{C}F_\LS(t)\Pi_\mathcal{C} ,\rho_\S]$ corresponds to a unitary evolution inside $\mathcal{C}$. Therefore, we can cancel its effect  up to $\mathcal{O}(\lambda^3)$, by applying the unitary  
$U^\dag_\LS(t)\equiv e^{i\lambda^2  \Pi_\mathcal{C} F_\LS(t) \Pi_\mathcal{C}}$. Indeed, since this unitary preserves the code subspace, instead of applying an active unitary, we can take its effect into account when we measure or interact with the system later (or, we can exploit it to implement non-trivial gates).   After applying this unitary, the restriction of state to $\mathcal{C}$ is equal to
$\rho_\S-\frac{\lambda^2}{2} \{\rho_\S, A(t)\}+\mathcal{O}(\lambda^3)$.  Therefore,  for any pure initial state $\rho_\S=|\psi\rangle\langle\psi|$ in $\mathcal{C}$, 
\beq\label{req}
F^2(|\psi\rangle, U_\LS^\dag(t) \rho_\S(t) U_\LS(t))=1-\lambda^2 \langle\psi|A(t)|\psi\rangle+\mathcal{O}(\lambda^3)\ .
\eeq
Using the joint-concavity of fidelity \cite{nielsen2000quantum} together with the linearity of time evolution, this yields a lower bound on $ F^2(\rho_\S, U_\LS^\dag(t)\rho_\S(t) U_\LS(t))$ for arbitrary initial mixed state $\rho_\S$ in  $\mathcal{C}$. Furthermore, note that by tracing over both sides of Eq.(\ref{eqasd}) we find that the right-hand side of Eq.(\ref{req})  is also equal to  $\Tr(\rho_\S(t)\Pi_C)$, the probability that at time $t$ system S is  inside $\mathcal{C}$.  Finally, by expressing the coefficients $b_{\mu\mu}(t)$, which determine the operator $A(t)= \sum_{\mu\ge E_\text{gap}} b_{\mu,\mu}(t) \ \Pi_\mathcal{C}  S_{\mu}S^\dag_\mu \Pi_\mathcal{C}$, in terms of $
p^\E_{B}(\omega)= \int dt\ e^{i\omega t}\ \Tr\big(\rho_\B\ e^{i H_\B t}B e^{-i H_\B t} B^\dag \big)\ $,  the \emph{Power Spectral Density} (PSD) of the environment operator $B$, we arrive at our first result:
\begin{theorem}\label{mainlemma10}
Assuming the error detection condition Eq.(\ref{cond})  holds, and neglecting the terms of $\mathcal{O}(\lambda^3)$ and higher, we find 
\begin{align}\label{mainlemma}
\underset{\rho_S\in \mathcal{S}(\mathcal{C})}{\min} &F^2(\rho_\S, U_\LS^\dag(t)\rho_\S(t) U_\LS(t))=\underset{\rho_S\in \mathcal{S}(\mathcal{C})}{\min} \Tr(\rho_\S(t)\Pi_C) \nonumber\\  &=1-\lambda^2\Big\| \sum_{E_n\ge E_\text{gap}} b_{n}(t)\ \Pi_\mathcal{C}  S \Pi_{E_n} S\Pi_\mathcal{C}\Big\|\ ,
\end{align}
where $U^\dag_\LS(t)\equiv e^{i\lambda^2  \Pi_\mathcal{C} F_\LS(t) \Pi_\mathcal{C}}$ is a unitary inside $\mathcal{C}$, $\mathcal{S}(\mathcal{C})$ is the set of states in the code subspace, $b_{n}(t)= \frac{1}{2\pi}\int d\omega\ p_B^\E(\omega)\left[\frac{\sin (\omega+E_n)t/2}{(\omega+E_n)/2}\right]^2$,  and $\|\cdot\|$ is the operator norm.
\end{theorem}
Theorem \ref{mainlemma10}  provides a general framework for understanding how error suppression works in the perturbative regime (up to $\mathcal{O}(\lambda^2)$). It implies that the fidelity loss happens either because of leakage out of $\mathcal{C}$, or unitary evolution inside $\mathcal{C}$. In other words,  if the system stays inside the code subspace then it  just evolves unitarily, and does not decohere. Note that this simple picture is not true in general, when the error detection condition does not hold. 

Another corollary of theorem \ref{mainlemma10} is that if the error detection condition holds, then the only relevant property of the environment which determines the speed of decoherence is the PSD of the environment operator. Now suppose we increase the gap by $\Delta E>0$, i.e. we add a penalty term $\Delta E(I-\Pi_\mathcal{C})$ to the Hamiltonian, whereby penalizing all states outside  $\mathcal{C}$ by an extra $\Delta E$. Then, theorem \ref{mainlemma10}  implies that this is equivalent to  replacing the PSD $p^\E_B(\omega)$ in the integral that yields coefficients $b_n(t)$ with  $p^\E_B(\omega-\Delta E)$. In other words, the  effect of adding this energy penalty is exactly equivalent to a shift of PSD by $\Delta E$.  Note that, in general, the PSD $p^\E_B(\omega)$  is negligible at  large frequencies, and therefore, by the above argument we find that leakage and decoherence are suppressed in the large $\Delta E$ limit. 
 
 

Theorem \ref{mainlemma10} is derived under the sole assumption that (i) the error detection condition holds, and (ii) the coupling is weak, $\lambda\ll 1$. In the large $t$ limit, and under extra  assumptions about decay of  correlations in the environment,  the coefficients  $b_n(t)$  can be approximated by $t p^\E_B(-E_n)$ (This is basically the regime where the rates of transitions are given by  the Fermi Golden Rule, and the Born-Markov approximation can be applied).  However,  these  assumptions do not generally hold in many cases of interest, e.g. where the memory of environment and the corresponding non-Markovian effects are non-negligible. Also, in the case of finite spin systems,  where  PSD is sum of delta functions, these approximations are not valid, whereas  Eq.(\ref{mainlemma}) remains true.  Next, we focus on the case of environments with local and bounded interactions, and present a bound on the PSD of local observables at high  frequencies. 

\textit{Locality of environment}.--- 
The sole assumption that interactions in a many-body system are local and bounded has far-reaching consequences. A well-known example is the finite speed of propagation of information, i.e. the Lieb-Robinson bound \cite{LiebRobinson, hastings2010locality}, which itself is used to prove many other general properties of these systems (See e.g. \cite{Hastings_area, bravyi2006lieb, hastings2010locality}). Here we explore another manifestation of locality, namely the fact that although  a many-body system  has arbitrary large frequencies, from the point of view of a local observer, who observes or interacts with the system locally,  the effect of the large frequencies are negligible, and the relevant frequencies are mainly determined by the strength of local interactions. We formalize this intuition in terms of $p^\E_B(\omega)$, the PSD  of a local operator $B$ of a many-body system with Hamiltonian $H_\B$.  

In the following, we assume $H_\B$ is local and bounded  around the support of $B$.  This means that it can be decomposed as $H_\B= \sum_i h_i +H_\B^{\text{U}}$, where $H_\B^{\text{U}}$ is completely unrestricted but it acts far from the support of $B$, and $h_i$ are local and bounded interactions in the neighborhood around the support of $B$, such that:  
\textbf{(i)} The strength of local interactions $h_i$ are bounded by $J_\text{max}>0$, i.e. $\|h_i\| \le J_\text{max}\ $, whereas $H_\B^{\text{U}}$ can be unbounded. \textbf{(ii)}  Interactions $h_i$ are $k-$local, i.e. each  acts non-trivially on, at most, $k$ sites in the system.  Also, the number of distinct  $h_i$ which act non-trivially on a single site is, at most, $r$. Therefore, $r$ and $k$ characterize the locality of $\sum_i h_i$, the restricted part of $H_\B$. For instance, on a $d$-dimensional rectangular lattice with nearest neighbor interactions $r=2d$ and $k=2$. Again, note that $H_\B^{\text{U}}$ can be non-local. \textbf{(iii)} $R_B$, the number of interactions $h_i$  which do not commute with $B$ is finite. Two operators do not commute with each other only if they have overlapping supports. Hence, roughly speaking, $R_B$ characterizes the \emph{non-locality} of operator $B$.  \textbf{(iv)} The supports of the unrestricted part $H_\B^{\text{U}}$  and operator $B$ are non-overlapping. Let $l>0$ be the length of the shortest path between the supports of $B$ and  $H_\B^{\text{U}}$ on the interaction graph. That is, $l$ is  the minimum number of interactions $h_i$ required to connect the support of these two operators.   

 We phrase our  result in terms of the \emph{cumulative PSD} of operator $B$,  defined as $P_B^\E(\omega)=\int_{|\mu|\ge|\omega|} d\mu\  p_B^{\E}(\mu)$, that is the total power in the positive and negative  frequencies larger than or equal to $|\omega|$. Then, 
using techniques similar to those used in the proof of the Lieb-Robinson bound \cite{LiebRobinson, hastings2010locality}, and the result of \cite{arad2014connecting}, in the supplementary material we prove  that
\begin{align} \label{main200}
P^\E_B(\omega)&\le \|B\| \sqrt{2\pi P^\E_B(0^+)} \times \frac{2^\frac{R_B}{rk}+1}{\text{Exp}_l\big(\frac{|\omega|}{8 J_\text{max} r k} \big) } \  , 
\end{align}
where $P^\E_{{B}}(0^+)=\int_{|\mu|>0} d\mu\ p^\E_B(\mu) \le 2\pi \|B\|^2$ can be interpreted as the total AC power of the fluctuations of $B$, and is zero if $B$ is conserved. Furthermore, $\text{Exp}_l(x)\equiv \sum_{k=0}^l {x^k}/{k!}$  is the truncated Taylor series of the exponential function $e^x$ at the $l$-th order, which up to a small multiplicative error, can be approximated by $e^{x}$ for  $0\le x\ll l$. Hence, Eq. (\ref{main200}) implies that in the regime $|\omega|\le l (J_\text{max} r k)$, the PSD should decay, at least, exponentially fast with $|\omega|$. The special case where $H_\B^{U}=0$, i.e. the case where Hamiltonian $H_\B$ is local and bounded everywhere throughout the system, corresponds to $l=\infty$, in which case the right-hand side of Eq.(\ref{main200}) becomes   $ e^{-|\omega|/(8 J_\text{max} r k)}$ multiplied by a frequency-independent term.

Note that non-locality and unboundedness of Hamiltonian $H_\B$ at distant points do not affect our bound drastically. 
Also, note that the right-hand side of bound  (\ref{main200}) increases exponentially fast with $R_B$, the number of local interactions which do not commute with $B$. For a typical non-local observable this quantity will be large, which is consistent with the fact that 
a non-local observable can see large frequencies of the system. See  \cite{prep} for further discussion about applications of bound Eq.(\ref{main200}).

\textit{Exponential suppression of errors}.--- Combining theorem \ref{mainlemma10} and bound (\ref{main200}) we can find an upper bound on the fidelity loss for environments with local and bounded interactions: we decompose the integral $b_n(t)= (2\pi)^{-1}\int d\omega\ p_B^\E(\omega)\left[\frac{\sin (\omega+E_n)t/2}{(\omega+E_n)/2}\right]^2$ in theorem \ref{mainlemma10} as the sum of two integrals over the intervals $(-\infty,-E_\text{gap}/2)$ and $[-E_\text{gap}/2,\infty)$. In the second interval, for  $E_n\ge E_\text{gap}$, which are the only relevant energies in Eq.(\ref{mainlemma}), function $[\frac{\sin (\omega+E_n)t/2}{(\omega+E_n)/2}]^2$  is bounded by $16/E^2_\text{gap}$. On the other hand, in the first interval, where this function could be as large as $t^2$, we use our upper bound on the cumulative PSD in Eq.(\ref{main200}). Then, as we show in the Supplementary Material, we arrive at 
\begin{theorem}\label{Thm11}
Suppose the interaction $\lambda H_\I$ satisfies the error detection condition in Eq.(\ref{cond}), and the environment Hamiltonian $H_\B$ satisfies conditions \textbf{(i-iv)}. Then 
\begin{align}\label{mainmain}
&1-\text{F}^2\big(U^\dag_\LS(t) \rho_\S(t) U_\LS(t),\rho_\S\big)\le\nonumber  \\ &\ \ \ \ \lambda^2\frac{16\|\Pi_\mathcal{C} H^2_\I  \Pi_\mathcal{C} \|}{E^2_\text{gap}}+  \lambda^2 \frac{t^2 \|\Pi_\mathcal{C} H^2_\I  \Pi_\mathcal{C} \|}{Q(E_\text{gap})}
 +\mathcal{O}(\lambda^3)\ ,
\end{align}
where the suppression factor $Q(E_\text{gap})$ is equal to $\text{Exp}_l\big(\frac{E_\text{gap}}{16 J_\text{max} r k} \big)\times (2^{\frac{R_B}{rk}}+1)^{-1}$.\end{theorem}
Note that the special case where the interactions  are local and bounded everywhere  throughout the environment, i.e. $H_\B=\sum h_i$, corresponds to $l\rightarrow\infty$. In this case the suppression factor is $Q(E_\text{gap})=e^{E_\text{gap}/(16 r k J_\text{max})}\times(2^{\frac{R_B}{rk}}+1)^{-1}$, and thus decoherence is suppressed by a factor which grows exponentially fast with the ratio $E_\text{gap}/(16 r k J_\text{max})$. This can be compared with the recent result of Bookatz et al \cite{bookatz2015error}, $1-\text{F}^2\big(\rho_\S(t),\rho_\S\big)\le \lambda^2 \|H_\I \|^2E^{-2}_\text{gap} [J_\text{max}t\mathcal{O}(1)+\mathcal{O}(1)]^2+\mathcal{O}(\lambda^3)$, which is obtained assuming  $k,r$ and $R_B$ are $\mathcal{O}(1)$.


Note that the term $\lambda^2\|\Pi_\mathcal{C} H^2_\I  \Pi_\mathcal{C} \|/E^2_\text{gap}$ in the right-hand side of Eq.(\ref{mainmain}) corresponds to the leakage due to the effect of the coupling Hamiltonian $\lambda H_\I$ itself, and it exists even for $H_\B=0$. 
Since this term is time-independent and  $\lambda\ll 1$, its effect  remains insignificant.   The second term, on the other hand, corresponds to the errors due to the environment Hamiltonian, and it vanishes for $H_\B=0$. This term grows with time, and  has the main contribution in decoherence in the weak coupling limit.  Therefore, our result on the exponential growth of suppression factor $Q(E_\text{gap})$, guarantees that decoherence remains small for a time which increases exponentially with $E_\text{gap}$. In particular,  Eq.(\ref{mainmain}) implies that  for time $t\approx \sqrt{Q(E_\text{gap})}/E_\text{gap}$ the total fidelity loss remains of the same order of the fidelity loss due to effect of the coupling Hamiltonian $\lambda H_\I$ itself, i.e. $\approx \lambda^2\|\Pi_\mathcal{C} H^2_\I  \Pi_\mathcal{C} \|/E^2_\text{gap}$, and hence is negligible in the weak coupling limit. Note that this result holds regardless of the size or temperature of the environment.

 We conclude that for this model of environment the effectiveness of error suppression with finite energy penalties is mainly determined by two properties of the environment: (i) the strength of the local interactions in the neighborhood around the region which interacts with the system, quantified by $J_\text{max}$ and (ii) the locality of the interactions in this neighborhood, captured by parameters $r$ and $k$. 
 
 

\textit{Relaxation of spin systems}.--- Theorem \ref{Thm11} can also be used to study relaxation of spin systems  with non-uniform interactions. Note that for a non-degenerate  ground subspace $\mathcal{C}$ the error detection condition always holds trivially, and the left hand side of Eq.(\ref{mainmain}) is simply the probability of leaving the ground state. Then, in the cases where the interactions are strong in one region and weak in the neighborhood around that region, theorem \ref{Thm11} can be applied to find a bound on the relaxation time, which is stronger than the bound set by the quantum speed limits \cite{marvian2015quantum}.

\textit{Discussion}.---Any approach for protecting quantum information has its own limitations, and is applicable only under certain assumptions about the nature of noise and the available resources. Our result on the exponential suppression of errors using energy penalty provides a strong evidence for the usefulness of this method for suppressing errors from certain types of environments, namely those which are bounded and local in a neighborhood around the region which interacts with the system. We noticed that error suppression with gap penalty is much more effective if we take into account the  unitary evolution of the system inside the code subspace caused by the Lamb shift effect. In the context of error suppression for adiabatic quantum computation \cite{ PhysRevA.74.052322, bookatz2015error},  this means that the adiabatic evolution inside the code subspace is effectively governed by a \emph{renormalized} Hamiltonian. Finally, we note that using the formal equivalence of dynamical decoupling \cite{Viola:99,Viola:98, PhysRevLett.100.160506}  and error suppression with energy penalty, shown in \cite{Young:13}, our approach  can also be adapted to study the effectiveness of this scheme of error suppression.



\textit{Acknowledgments}.--- 
I am grateful to Paolo Zanardi, Daniel Lidar and Adam Bookatz for reading this manuscript and providing useful comments. Also, I acknowledge helpful discussions with Edward Farhi and Seth Lloyd. This work was supported under grants ARO W911NF-12-1-0541 and NSF CCF-1254119.

\bibliography{ref}

\newpage

\appendix


\onecolumngrid

\newpage

\begin{center}
\Large{\textbf{Supplementary Material}}
\end{center}
\section{Second order approximation of the equation of motion (Proof of Eq.(2) in the paper)}

In this section we derive the reduced dynamics of the  system up to second order with respect to $\lambda$. We follow the approach of \cite{PhysRevA.40.4077,Alicki:02,Majenz:2013qw}. 

Consider a system $S$ with Hamiltonian $H_\S$, which interacts with an environment via the interaction $\lambda H_\I=\lambda S\otimes B$.  So the  total Hamiltonian at $t>0$ is  $H_\text{S}+\lambda {H}_\text{I}+H_\B $. Furthermore, consider the interaction picture, defined by the transformation $|\psi\rangle\rightarrow U^\dag_\S(t)\otimes U^\dag_\B(t) |\psi\rangle$ where $U_\S(t)=e^{-i H_\S t}$  and $U_\B(t)=e^{-i H_\B t}$. In this frame the joint state of system and environment  is 
 \beq
\tilde{\rho}_\SB(t)=\Big[U^\dag_\S(t)\otimes U^\dag_\B(t)\Big] {\rho}_\SB(t) \Big[ U_\S(t)\otimes U_\B(t)\Big]\ ,
\eeq
and the equation of motion in this frame is
\beq
\frac{d}{dt}\tilde{\rho}_\SB(t)=-i\lambda  [\tilde{H_\I}(t), \tilde{\rho}_\SB(t)]\ ,
\eeq
where 
 \beq
\tilde{H_\I}(t)=\Big[U^\dag_\S(t)\otimes U^\dag_\B(t)\Big] H_\I \Big[ U_\S(t)\otimes U_\B(t)\Big]\ .
\eeq
It follows that up to the second order with respect to $\lambda$, $\tilde{\rho}_\SB(t)$ is given by
\begin{align}
\tilde{\rho}_\SB(t)=\rho_\SB(0)- i  \int_0^t ds\  \lambda [\tilde{H}_\I(s),\rho_\SB(0)]+ (-i  \lambda)^2   \int_0^t ds_1\ \int_0^{s_1} ds_2\ [\tilde{H}_\I(s_1), [\tilde{H}_\I(s_2),\rho_\SB(0)]] +\mathcal{O}(\lambda^3) \ .
\end{align}
Then, up to the second order with respect to $\lambda$, the reduced state of system $\tilde{\rho}_\S(t)=\Tr_\B(\tilde{\rho}_\SB(t))$ is given by 
\begin{align}
\tilde{\rho}_\S(t)=\rho_\S- i  \int_0^t ds\  \lambda \Tr_\B\Big( [\tilde{H}_\I(s),\rho_\SB(0)]\Big)+ (-i\lambda)^2   \int_0^t ds_1\ \int_0^{s_1} ds_2\ \Tr_\B\Big( [\tilde{H}_\I(s_1), [\tilde{H}_\I(s_2),\rho_\SB(0)]] \Big)+\mathcal{O}(\lambda^3) \ .
\end{align}
Next, assume the initial joint state of system and environment  is uncorrelated, i.e. $\rho_\SB(0)=\rho_\S\otimes \rho_\B$. Furthermore, assume the environment is initially in equilibrium, i.e. $[\rho_\B, H_\B]=0$. Define $H^{(1)}_\I  =   \Tr_\B\Big(H_\I \rho_\B\Big)$ and 
\beq
\tilde{H}^{(1)}_\I(t)  =  U^\dag_\S(t)  \Tr_\B\Big(H_\I \rho_\B\Big) U_\S(t)\ .
\eeq
Then
\begin{align}\label{ff123}
\tilde{\rho}_\S(t)=\rho_\S- i  \lambda \int_0^t ds\  [\tilde{H}^{(1)}_\I(s)  ,\rho_\S]  + (-i\lambda)^2   \int_0^t ds_1\ \int_0^{s_1} ds_2\ \Tr_\B\Big( [\tilde{H}_\I(s_1), [\tilde{H}_\I(s_2),\rho_\SB(0)]] \Big)+\mathcal{O}(\lambda^3) \ .
\end{align}
Next, we focus on the term of $\mathcal{O}(\lambda^2)$. Consider the decomposition  $S=\sum_{\mu\in \Omega} S_\mu$, where $S_\mu$ is the component of the system operator $S$ in frequency $\mu$ with respect to $H_\S$, i.e. $S_\mu=\sum_n \Pi_{E_n} S \Pi_{E_n +\mu}$. (Recall that $H_\S=\sum_n E_n\ \Pi_{E_n}$ is the spectral decomposition of the system Hamiltonian,  and $\Omega$ is the set of corresponding Bohr frequencies, i.e. the set of all energy difference $E_n-E_m$ in the system). This decomposition implies 
\begin{align}
\tilde{H}_\I(s)= [U^\dag_\S(s)\otimes U^\dag_\B(s)] H_\I [U_\S(s)\otimes U_\B(s)]=  \sum_{\mu\in\Omega} e^{-i\mu s} S_\mu\otimes\tilde{B}(s)=\sum_{\mu\in\Omega} e^{i\mu s} S^\dag_\mu\otimes\tilde{B}(s)  ,
\end{align}
where we have used the fact that $S_{-\mu}=S^\dag_{\mu}$.  Using this we find that the term of order $\lambda^2$ in Eq.(\ref{ff123})  is equal to
\begin{align}\label{eer1}
&-\lambda^2\sum_{\mu_1,\mu_2}\int_0^t ds_1\  \int_0^{s_1} ds_2\ \Big[ e^{i(-\mu_1 s_1+\mu_2 s_2)}\Tr(\tilde{B}(s_1) \tilde{B}(s_2) \rho_\B  )\   S_{\mu_1} S_{\mu_2}^\dag \rho_\S+ \text{H.C.} \Big]\nonumber\\ &\ \ \ \ \ \ \ \ +\lambda^2\sum_{\mu_1,\mu_2}\int_0^t ds_1\  \int_0^{s_1} ds_2\ \Big[ e^{i(-\mu_1 s_1+\mu_2 s_2)}\Tr(\tilde{B}(s_1)\rho_\B \tilde{B}(s_2) )\   S_{\mu_1}  \rho_\S S_{\mu_2}^\dag+ \text{H.C.} \Big]\ .
\end{align}
Define
\beq
\Gamma_{\mu_1 \mu_2}(t)= \int_0^t ds_1\  \int_0^{s_1} ds_2\ e^{i(-\mu_1 s_1+\mu_2 s_2)}\Tr(\tilde{B}(s_2-s_1) B \rho_\B  ) 
\eeq
Then the second line of Eq.(\ref{eer1}) reads as  
\bes\label{aws}
\begin{align}
\lambda^2\left [\sum_{\mu_1,\mu_2} \Gamma_{\mu_1\mu_2}(t)  S_{\mu_1}  \rho_\S S_{\mu_2}^\dag+ \sum_{\mu_1,\mu_2} \Gamma^\ast_{\mu_1\mu_2}(t)  S_{\mu_2}  \rho_\S S_{\mu_1}^\dag\right]&=\lambda^2\left [\sum_{\mu_1,\mu_2} \Gamma_{\mu_1\mu_2}(t)  S_{\mu_1}  \rho_\S S_{\mu_2}^\dag+ \sum_{\mu_2,\mu_1} \Gamma^\ast_{\mu_2\mu_1}(t)  S_{\mu_1}  \rho_\S S_{\mu_2}^\dag\right]\ \\ &= \lambda^2 \sum_{\mu_1,\mu_2} \left[\Gamma_{\mu_1\mu_2}(t)+\Gamma^\ast_{\mu_2\mu_1}(t)\right]  S_{\mu_1}  \rho_\S S_{\mu_2}^\dag\ ,
\end{align}
\ees
where the first equality is obtained by exchanging labels $\mu_1$ and $\mu_2$ in the second summation.
Next we note that
\bes
\begin{align}
\Gamma_{\mu_1 \mu_2}(t)+\Gamma^\ast_{\mu_2 \mu_1}(t)&= \int_0^t ds_1\  \int_0^{s_1} ds_2\ e^{-i(\mu_1 s_1-\mu_2 s_2)}\Tr(\tilde{B}(s_2-s_1) B \rho_\B  )+\int_0^t ds_1\  \int_0^{s_1} ds_2\ e^{i(\mu_2 s_1-\mu_1 s_2)}\Tr(\tilde{B}(s_1-s_2) B \rho_\B  ) \\ &
= \int_0^t ds_1\  \int_0^{s_1} ds_2\ e^{-i(\mu_1 s_1-\mu_2 s_2)}\Tr(\tilde{B}(s_2-s_1) B \rho_\B  )+\int_0^t ds_2\  \int_0^{s_2} ds_1\ e^{i(\mu_2 s_2-\mu_1 s_1)}\Tr(\tilde{B}(s_2-s_1) B \rho_\B  ) \\ &
= \int_0^t ds_1\  \int_0^{t} ds_2\ e^{-i(\mu_1 s_1-\mu_2 s_2)}\Tr(\tilde{B}(s_2-s_1) B \rho_\B  ) \\ &
= \int_0^t ds_1\  \int_0^{t} ds_2\ e^{-i(\mu_1 s_1-\mu_2 s_2)}\Tr^\ast(\tilde{B}(s_1-s_2) B \rho_\B  )\ ,
\end{align}
\ees
where  in the first line we have used $\Tr^\ast(\tilde{B}(s_2-s_1) B \rho_\B )=\Tr(\tilde{B}(s_1-s_2) B \rho_\B )$, and to get the second line we have exchanged $s_1$ and $s_2$ in the second  term. Then, using the definition 
\beq
b_{\mu_1\mu_2}(t)=\int_0^t\int_0^t ds_1 ds_2 \ e^{i(\mu_2 s_2-\mu_1 s_1)}  \Tr(\rho_\B \tilde{B}(s_1-s_2)B)\ ,
\eeq
we find
\bes\label{qwer}
\begin{align}
\Gamma_{\mu_1 \mu_2}(t)+\Gamma^\ast_{\mu_2 \mu_1}(t)&
= \int_0^t ds_1\  \int_0^{t} ds_2\ e^{-i(\mu_1 s_1-\mu_2 s_2)}\Tr^\ast(\tilde{B}(s_1-s_2) B \rho_\B  )
\\ &=b^\ast_{-\mu_1-\mu_2}(t)\ .
\end{align}
\ees
This together with Eq.(\ref{aws}) implies that  the second line of Eq.(\ref{eer1}) is equal to 
\bes\label{asad}
\begin{align}
 \lambda^2 \sum_{\mu_1,\mu_2} b^\ast_{-\mu_1-\mu_2}(t)   S_{\mu_1} \rho_\S S_{\mu_2}^\dag&=\lambda^2 \sum_{\mu_1,\mu_2} b^\ast_{\mu_1\mu_2}(t)   S^\dag_{\mu_1} \rho_\S S_{\mu_2}\\ &=\lambda^2 \sum_{\mu_1,\mu_2} b_{\mu_2\mu_1}(t)   S^\dag_{\mu_1} \rho_\S S_{\mu_2}\ ,
\end{align}
\ees
where to get the first equality we have used the fact that $S_{-\mu}=S^\dag_{\mu}$, and we have used the fact that $ b^\ast_{\mu_1\mu_2}(t) =b_{\mu_2\mu_1}(t)$.

Next, we focus on the first line of Eq.(\ref{eer1}). 
Using the fact that $\int_0^t\int_0^{s_1} ds_2\  e^{i(\mu_1 s_1-\mu_2 s_2)}\Tr(\tilde{B}(s_1) \tilde{B}(s_2) \rho_\B  ) =\Gamma^\ast_{\mu_1,\mu_2}(t)$, we find that this term can be written as
\beq
-\lambda^2\left[\sum_{\mu_1,\mu_2}   \Gamma^\ast_{-\mu_1,-\mu_2}(t)  S_{\mu_1} S_{\mu_2}^\dag \rho_\S +\sum_{\mu_1,\mu_2}   \Gamma_{-\mu_1,-\mu_2}(t)   \rho_\S S_{\mu_2}S^\dag_{\mu_1}   \right]\\
=-\lambda^2\left[\sum_{\mu_1,\mu_2}   \Gamma^\ast_{\mu_1,\mu_2}(t) S^\dag_{\mu_1} S_{\mu_2}   \rho_\S +\sum_{\mu_1,\mu_2}   \Gamma_{\mu_1,\mu_2}(t)   \rho_\S S^\dag_{\mu_2}  S_{\mu_1} \right]\ ,
\eeq
where we have used $S_{-\mu}=S^\dag_{\mu}$ in the second term. Then, we note that
\bes
\begin{align}
&-\lambda^2\left[\sum_{\mu_1,\mu_2}   \Gamma^\ast_{\mu_1,\mu_2}(t) S^\dag_{\mu_1} S_{\mu_2}   \rho_\S +\sum_{\mu_1,\mu_2}   \Gamma_{\mu_1,\mu_2}(t)   \rho_\S  S^\dag_{\mu_2} S_{\mu_1} \right]\\ &=-\lambda^2 \sum_{\mu_1,\mu_2}   \Gamma^\ast_{\mu_1,\mu_2}(t) \frac{1}{2}\Big( [S^\dag_{\mu_1} S_{\mu_2} , \rho_\S]+\{S^\dag_{\mu_1} S_{\mu_2} , \rho_\S\}\Big) -\lambda^2\sum_{\mu_1,\mu_2}   \Gamma_{\mu_1,\mu_2}(t)   \frac{1}{2}\Big( [\rho_\S,  S^\dag_{\mu_2} S_{\mu_1}]+\{ S^\dag_{\mu_2} S_{\mu_1}, \rho_\S\}\Big)\\ &=-\lambda^2 \sum_{\mu_1,\mu_2}   \Gamma^\ast_{\mu_1,\mu_2}(t) \frac{1}{2}\Big( [S^\dag_{\mu_1} S_{\mu_2} , \rho_\S]+\{ S^\dag_{\mu_1} S_{\mu_2}, \rho_\S\}\Big) -\lambda^2\sum_{\mu_1,\mu_2}   \Gamma_{\mu_2,\mu_1}(t)   \frac{1}{2}\Big( [\rho_\S, S^\dag_{\mu_1} S_{\mu_2} ]+\{ S^\dag_{\mu_1}S_{\mu_2}, \rho_\S\}\Big)\\ &=-\lambda^2 \sum_{\mu_1,\mu_2}    \frac{\Gamma^\ast_{\mu_1,\mu_2}(t)+ \Gamma_{\mu_2,\mu_1}(t)}{2} \{S^\dag_{\mu_1}S_{\mu_2} , \rho_\S\} -\lambda^2 \sum_{\mu_1,\mu_2}    \frac{\Gamma^\ast_{\mu_1,\mu_2}(t)- \Gamma_{\mu_2,\mu_1}(t)}{2} [ S^\dag_{\mu_1}S_{\mu_2}, \rho_\S]\ ,
\end{align}
\ees
where to get the third line we have exchanged $\mu_1$ and $\mu_2$ in the second term. Then, using Eq.(\ref{qwer}), we have $\Gamma_{\mu_2 \mu_1}(t)+\Gamma^\ast_{\mu_1 \mu_2}(t)=b^\ast_{-\mu_2-\mu_1}(t)$. This implies that the first line in Eq.(\ref{eer1}) can be rewritten as
\begin{align}
&-\lambda^2 \sum_{\mu_1,\mu_2}    \frac{b^\ast_{-\mu_2-\mu_1}(t)}{2} \{S^\dag_{\mu_1} S_{\mu_2} , \rho_\S\} -\lambda^2 \sum_{\mu_1,\mu_2}    \frac{\Gamma^\ast_{\mu_1,\mu_2}(t)- \Gamma_{\mu_2,\mu_1}(t)}{2} [S^\dag_{\mu_1} S_{\mu_2} , \rho_\S]
\\ &=-\lambda^2 \sum_{\mu_1,\mu_2}    \frac{b^\ast_{\mu_2\mu_1}(t)}{2} \{S_{\mu_1} S^\dag_{\mu_2} , \rho_\S\} -\lambda^2 \sum_{\mu_1,\mu_2}    \frac{\Gamma^\ast_{\mu_1,\mu_2}(t)- \Gamma_{\mu_2,\mu_1}(t)}{2} [S^\dag_{\mu_1} S_{\mu_2} , \rho_\S]
\\ &=-\lambda^2 \sum_{\mu_1,\mu_2}    \frac{b_{\mu_1\mu_2}(t)}{2} \{ S_{\mu_1} S^\dag_{\mu_2} , \rho_\S\} -\lambda^2 \sum_{\mu_1,\mu_2}    \frac{\Gamma^\ast_{\mu_1,\mu_2}(t)- \Gamma_{\mu_2,\mu_1}(t)}{2} [S^\dag_{\mu_1} S_{\mu_2} , \rho_\S]
\end{align}
where to get the third line we have used the fact that $b^\ast_{\mu_2\mu_1}(t)=b_{\mu_1\mu_2}(t)$. Using this together with Eq.(\ref{asad}) which gives the second line of Eq.(\ref{eer1}),  we find
\begin{align}
\tilde{\rho}_\S(t)&=\rho_\S- i  \lambda \int_0^t ds\  [\tilde{H}^{(1)}_\I(s)  ,\rho_\S] +    \lambda^2 \sum_{\mu_1,\mu_2} b_{\mu_2\mu_1}(t)   S^\dag_{\mu_1} \rho_\S S_{\mu_2}\ -\lambda^2 \sum_{\mu_1,\mu_2}    \frac{b_{\mu_1\mu_2}(t)}{2} \{S_{\mu_1  }S^\dag_{\mu_2}, \rho_\S\} - [i \lambda^2 F_\LS(t), \rho_\S] \\
&=\rho_\S+ \lambda^2 \sum_{\mu_1,\mu_2} b_{\mu_2\mu_1}(t)   S^\dag_{\mu_1} \rho_\S S_{\mu_2}\ -\lambda^2 \sum_{\mu_1,\mu_2}    \frac{b_{\mu_2\mu_1}}{2} \{S_{\mu_2} S^\dag_{\mu_1}, \rho_\S\} - i \Big[ \lambda^2 F_\LS(t)+\lambda \int_0^t ds \tilde{H}^{(1)}_\I(s) , \rho_\S\Big] \ ,
\end{align}
where
\beq
\lambda^2 F_\LS(t)=\lambda^2 \sum_{\mu,\mu'} h_{\mu_1\mu_2}(t) S^\dag_{\mu_1} S_{\mu_2}
\eeq
is the Lamb shift Hamiltonian, and 
\beq
h_{\mu_1\mu_2}(t)=\frac{\Gamma^\ast_{\mu_1,\mu_2}(t)- \Gamma_{\mu_2,\mu_1}(t)}{2i}\ .
\eeq

\newpage
\section{Equilibrium power spectral density (proof of Eq.(9) in the paper)}

For completeness we first review the assumptions and the main result. Let $\rho_\E$ be an equilibrium state for Hamiltonian $H$, i.e. $[\rho_\E,H]=0$. Let $ p_A^\E(\omega)= \int dt\ e^{i\omega t} \Tr(A^\dag \rho_\E A(t))$ be the equilibrium power spectral density (PSD) of operator $A$, where $A(t)=e^{i Ht} A e^{-i Ht}$. Let $H=\sum_k  E_k \Pi_{E_k}$ be the spectral decomposition of $H$, and 
\beq
A_\text{diag}=\lim_{T\rightarrow \infty}\frac{1}{T}\int_0^T dt\ A(t)=\sum_{k} \Pi_{E_k} A \Pi_{E_k} \ ,
\eeq
be the component of $A$ which is diagonal in the Hamiltonian eigenbasis. 

\begin{theorem}\label{Thm:app}
Let $A$ be an arbitrary linear operator on the Hilbert space of the system. Suppose the system Hamiltonian $H$ can be decomposed as $H= \sum_i h_i +H^{\text{U}}$ where 

\noindent\textbf{(i)} The support of the unrestricted part $H^{\text{U}}$ and the support of $A$ are non-overlapping.   Furthermore, the number of interactions $h_i$  which do not commute with operator $A$ is  $R_A$. 

\noindent\textbf{(i)} The maximum strength of local interactions $h_i$ is bounded by $J_\text{max}$, i.e. $\|h_i\| \le J_\text{max}\ $. 

\noindent\textbf{(iii)}   All $h_i$ are $k-$local, i.e. each of them acts non-trivially on at most $k$ sites in the system.  Furthermore, the number of distinct  $h_i$ which act non-trivially on a single site in the system is at most $r$. 

\noindent\textbf{(iv)} The length of the shortest path between the support of observable $A$ and the support of  $H^{\text{U}}$ on the \emph{interaction graph} induced by $H= \sum_i h_i +H^{\text{U}}$ is $l$. In other words, $l$ is the minimum number of interactions $h_i$ which are needed to connect the supports of $A$ and $H^{\text{U}}$.

Then for $1\le n\le l$ it holds that
\begin{align}\label{main_app}
\int d\omega \ |\omega|^n\ p^\E_A(\omega) &\le   2\pi  \sqrt{\Tr(\rho_\text{eq} A A^\dag)-\Tr( \rho_\text{eq} A_\text{diag} A_\text{diag}^\dag)}\times \|A\|\times 2^\frac{R_A}{rk}(4rk J_\text{max})^n n! \  .
\end{align}
Furthermore,
\begin{align}\label{main200_app}
P^\E_A(\omega)\equiv \int_{|\mu|\ge |\omega|} d\mu \  p^\E_A(\mu)  &\le   2\pi  \sqrt{\Tr(\rho_\text{eq} A A^\dag)-\Tr( \rho_\text{eq} A_\text{diag} A_\text{diag}^\dag)}\times   \|A\|\times \frac{2^\frac{R_A}{rk}+1}{\text{Exp}_l(\frac{|\omega|}{8rk J_\text{max}})} \  ,
\end{align}
where $\text{Exp}_l(x)\equiv \sum_{k=0}^l {x^k}/{k!}$  is the truncated Taylor series of the exponential function $e^x$ at the $l$-th order.
\end{theorem}
\noindent\textbf{Remark:} Note that $\Tr( \rho_\text{eq} A_\text{diag} A_\text{diag}^\dag)$ is indeed the time average of $ \Tr(  \rho_\E A(t) A^\dag)$, i.e. 
\beq
\Tr( \rho_\text{eq} A_\text{diag} A_\text{diag}^\dag)=\lim_{T\rightarrow \infty} \frac{1}{T} \int_{0}^T dt\  \Tr(  \rho_\E A(t) A^\dag)\ ,
\eeq
which follows from the fact that $\rho_\E$ is diagonal in the Hamiltonian eigenbasis. Then, using the fact that $\int dt\ e^{i\omega t}=2\pi\delta(\omega)$, we find that the  PSD function $p^\E_{A}(\omega)$ at $\omega=0$ is equal to $2\pi \Tr( \rho_\text{eq} A_\text{diag} A_\text{diag}^\dag) \delta(\omega)$. 
Furthermore, using $\int d\omega p(\omega)=2\pi  \Tr(\rho_\text{eq} A A^\dag)$, it follows that
\begin{align}\label{qrty}
 \Tr(\rho_\text{eq} A A^\dag)-\Tr( \rho_\text{eq} A_\text{diag} A_\text{diag}^\dag) &=
 \frac{1}{2\pi}\int_{|\omega|>0}  d\omega\ p^\E_{A}(\omega)=\frac{1}{2\pi} P^\E_{{A}}(0^+) \ . 
\end{align}
The quantity $P^\E_{{A}}(0^+)$ can be interpreted as the total  \emph{AC power} of fluctuations of operator $A$, i.e. the difference between the \emph{total power} $P^\E_{{A}}(0)= \int d\omega\ p^\E_{A}(\omega)$, and the \emph{DC power} $2\pi \Tr( \rho_\text{eq} A_\text{diag} A_\text{diag}^\dag)$. Eq.(9) in the paper follows from the above equation together with the bound in Eq.(\ref{main200_app}).

\subsection{Proof}
For any integer $n$  let ${A}^{(n)}(t)$ be  the $n$-th derivative of operator ${A}(t)=e^{i H t} A e^{-i H t}$, i.e.   
\beq
{A}^{(n)}(t)\equiv \frac{d^n }{dt^n} {A}(t)=i^n\times e^{i H t} \text{ad}^n_H(A) e^{-i H t} \ ,
\eeq
where for any operator $G$, $\text{ad}_G(X)$ denotes $[G,X]$. So,  at $t=0$ we have  $A^{(n)}(0)=i^n\times \text{ad}^n_H(A)$. For any arbitrary operator $B$ consider the inner product of $B$ and $A^{(n)}(0)$ given by
\begin{align}
\Tr\Big(  B^\dag \rho_\text{eq} A^{(n)}(0) \Big) &=i^n\times  \Tr\left(  B^\dag  \rho_\text{eq} \text{ad}^n_H(A) \right)\ .
\end{align}
Let $B_\text{diag}$ be the component of $B$ which is diagonal in the energy basis, i.e.
\beq
B_\text{diag}=\lim_{T\rightarrow \infty} \frac{1}{T} \int_0^T dt\   e^{i H t} B e^{-i H t}=\sum_{l} \Pi_{E_l} B\ \Pi_{E_l} \ ,
\eeq
where $H=\sum_l \Pi_{E_l} E_l$ is the spectral decomposition of $H$. 
Then, for $n\ge 1$ we find
\begin{align}\label{bb1}
\Tr\Big(  B^\dag \rho_\text{eq} A^{(n)}(0) \Big) &=i^n\times  \Tr\left(  B^\dag  \rho_\text{eq} \text{ad}^n_H(A) \right)=i^n\times \Tr\left(  [B^\dag-B_\text{diag}^\dag ] \ \rho_\text{eq} \text{ad}^n_H(A) \right)\ ,
\end{align}
where to get the second equality we have used the the cyclic property of the trace, together with the fact that $\rho_\text{eq} $ is a stationary state, and so it commutes with $H$, which implies $ \rho_\text{eq} B_\text{diag}^\dag $ commutes with $H$.  Then, using the Cauchy-Schwarz inequality we find that
\bes\label{erh}
\begin{align}
\big{|}\Tr\Big(  B^\dag \rho_\text{eq} A^{(n)}(0) \Big) \big{|}&=\Big|\Tr\left( [B^\dag-B^\dag_\text{diag} ]  \rho_\text{eq} \text{ad}^n_H(A)  \right)\Big|
\\ &\le  \sqrt{\Tr\Big((B^\dag-B^\dag_\text{diag})\rho_\text{eq} (B-B_\text{diag})  \Big)} \times \sqrt{\Tr\Big(  \text{ad}^n_H(A^\dag) \rho_\text{eq} \text{ad}^n_H(A) \Big) \times (-1)^n}\ 
\\ &\le  \sqrt{\Tr\Big((B^\dag-B^\dag_\text{diag})\rho_\text{eq} (B-B_\text{diag})  \Big)} \times \| \text{ad}^n_H(A)\| \ ,
\end{align}
\ees
where $\|\cdot\|$ is the infinity norm, i.e. the largest singular value of the operator. Using the fact that both $B_\text{diag}=\sum_{l} \Pi_{E_l} B\ \Pi_{E_l}$ and $\rho_\E$ commute with $H$, one can easily see that
\beq
\Tr\Big((B^\dag-B^\dag_\text{diag})\rho_\text{eq} (B-B_\text{diag})  \Big)=\Tr\Big(B^\dag \rho_\text{eq} B  \Big)-\Tr\Big(B_\text{diag}^\dag \rho_\text{eq} B_\text{diag}  \Big)\ .
\eeq
Putting this in Eq.(\ref{erh}) we find 
\begin{align}\label{bb12}
\big{|}\Tr\Big(  B^\dag \rho_\text{eq} A^{(n)}(0) \Big) \big{|} \le  \sqrt{\Tr(B^\dag \rho_\text{eq} B )-\Tr(B_\text{diag}^\dag \rho_\text{eq} B_\text{diag})} \times \| \text{ad}^n_H(A)\| \ ,
\end{align}
for all $n\ge 1$.

Next, consider the correlation function $\Tr\big(B^\dag \rho_\text{eq}  A(t) \big)$ and let $q_{AB}(\omega)=\int dt\ e^{i\omega t } \Tr\big(B^\dag \rho_\text{eq}  A(t) \big)$ be its Fourier transform. This means
\beq
\Tr\big( B^\dag \rho_\text{eq}  A(t)\big)=\frac{1}{2\pi} \int d\omega\ \ e^{-i\omega t}  \ q_{AB}(\omega)  \ .
\eeq
Taking the derivatives of both sides with respect to $t$ we find
\beq
\Tr\Big( B^\dag \rho_\text{eq} A^{(n)}(t)\Big)=\frac{1}{2\pi}\int d\omega\ \ (-i\omega)^n e^{-i\omega t}   \ q_{AB}(\omega)  \ .
\eeq
This together with Eq. (\ref{bb12})  implies that for all integer $n\ge 1$, it holds that
\begin{align}\label{asf}
\left|\int d\omega\ \ \omega^n  \ q_{AB}(\omega) \right|\le 2\pi  \sqrt{\Tr(B^\dag \rho_\text{eq} B )-\Tr(B_\text{diag}^\dag \rho_\text{eq} B_\text{diag})} \times  \Big\|   \text{ad}^n_H(A)  \Big\| \ .
\end{align}
This yields  a general bound on the moments of $ q_{AB}(\omega)$, the Fourier transform of correlation function  $\Tr\big(B^\dag \rho_\text{eq}  A(t) \big)$ for arbitrary (possibly non-local) Hamiltonian $H$ and operator $B$, and it could be of independent interest.

Next we use the assumptions about locality of operator $A$ and the locality and boundedness of Hamiltonian $H$ around the support of  $A$.  Consider the decomposition  of $H$ as  $H=H^R+H^U$, where $H^R=\sum_i h_i$ is the part of Hamiltonian which is bounded and local. By assumption,  the supports of $A$ and $H^U$ do not overlap with each other. This implies 
\beq
\text{ad}^1_H(A)=[H,A]=[H^U+H^R,A]=[H^R,A]=\text{ad}^1_{H^R}(A) \ .
\eeq
Then, we note that the support of $[H^R,A]$ is restricted to the union of the support of $A$ and the support of all local interactions $h_i$ which do not commute with $A$. By assumption $l$ is the minimum number of $h_i$ which are required to connect the support of $A$ to the support of $H^U$. Therefore,  if $l\ge 2$ then the supports of $[H,A]=[H^R,A]$ and $H^U$  do not overlap with each other, and therefore 
\beq
\text{ad}^2_H(A)=[H,[H,A]]=[H,[H^R,A]]=[H^R,[H^R,A]]=\text{ad}^2_{H^R}(A)\ .
\eeq
Repeating this argument we can easily see that for $n\le l$ we have 
\beq
\text{ad}^n_H(A)=\text{ad}^n_{H^R}(A)\ .
\eeq
Then, as we show in Sec.~\ref{sec:lem}, following \cite{arad2014connecting}, by counting the nonzero terms in $\text{ad}^n_{H^R}(A)$ and using the locality and boundedness of interactions $h_i$ we find that for $n\le l$,
\beq\label{counting}
\|\text{ad}^n_H(A)\|=\|\text{ad}^n_{H^R}(A)\|  \le (4 r k J_\text{max})^n \times n! \times 2^\frac{R_A}{rk}  \times \|A\| \ .
\eeq
Putting this in Eq.(\ref{asf}) we conclude that
\begin{align}\label{formula1}
\left|\int d\omega\ \ \omega^n  \ q_{AB}(\omega) \right| \le 2\pi  \sqrt{\Tr(B^\dag \rho_\text{eq} B )-\Tr(B_\text{diag}^\dag \rho_\text{eq} B_\text{diag})}  \times   (4rk J_\text{max})^n \times n! \times 2^\frac{R_A}{rk}  \times \|A\|  \ \ \ \ \ \ \ \ \ \ : 1\le n\le l\ .
\end{align}


In the following we show Eq.(\ref{main_app}) of theorem \ref{Thm:app},   follows from this bound. First, note that choosing $B=A$, Eq.(\ref{formula1}) implies that for any integer $n$ in the interval $1\le n\le l$,
\begin{align}\label{evencase}
\left|\int d\omega\ \ \omega^n  \ p^\E_{A}(\omega) \right| \le 2\pi  \sqrt{\Tr(A^\dag \rho_\text{eq} A )-\Tr(A_\text{diag}^\dag \rho_\text{eq} A_\text{diag})}  \times   (4rk J_\text{max})^n \times n! \times 2^\frac{R_A}{rk}  \times \|A\|  ,
\end{align}
where we have used the fact that $q_{AA}(\omega)=p_A^\E(\omega)$. This proves Eq.(\ref{main_app}) for even integers.

Next to prove  Eq.(\ref{main_app})  for odd integers, we choose $B$ to be 
\begin{align}\label{defA}
B=\overline{A}\equiv\sum_{k,l} \Pi_{E_k} A \Pi_{E_l} \times \text{Sign}(E_l-E_k) \ ,
\end{align}
where $H=\sum_l E_l \Pi_{E_l}$ is the spectral decomposition of Hamiltonian $H$, and $ \text{Sign}(x)$ is $+1$ for $x>0$, $-1$ for $x<0$ and $ \text{Sign}(0)=0$.  Equivalently, $\overline{A}$ can be defined in terms of the frequency decomposition of $A$ as $A=\sum_\omega \hat{A}(\omega)$ where
\beq
\hat{A}(\omega)\equiv\sum_l  \Pi_{E_l} A\Pi_{E_l+\omega}  .
\eeq 
Then $\overline{A}$ can be defined as, 
\beq\label{defA}
\overline{A}=\sum_\omega   \hat{A}(\omega) \text{Sign}(\omega)\ ,
\eeq 
which implies 
\beq\label{defA}
e^{i H t}\overline{A} e^{-i H t}=\sum_\omega   e^{-i\omega t} \hat{A}(\omega) \text{Sign}(\omega)\ ,
\eeq

Note that $\overline{A}$ can be a non-local operator, even though $A$ is local. Also, note that the discreteness of energies does not play any role in the following arguments; in the case of continuous spectrum we can simply replace the above summation with integral. 

Definition of $\overline{A}$ in Eq.(\ref{defA}), together with the fact that $\rho_\E$ commute with $H$  immediately implies that 
\bes\label{sign14}
\begin{align}
\Tr( \overline{A}^\dag\rho_\E  \overline{A} )&=\Tr( \left[\sum_\omega   \hat{A}^\dag(\omega) \text{Sign}(\omega)\right]\rho_\E  \left[\sum_{\omega'}   \hat{A}(\omega') \text{Sign}(\omega')\right])=\sum_\omega \Tr(  \hat{A}^\dag(\omega) \rho_\E  \hat{A}(\omega)) \times  \text{Sign}^2(\omega)\\ &=\sum_{\omega\neq 0} \Tr(  \hat{A}^\dag(\omega) \rho_\E  \hat{A}(\omega))\\ &= \Tr(\left[\sum_\omega   \hat{A}(\omega) \right]^\dag \rho_\text{eq} \left[\sum_\omega   \hat{A}(\omega) \right] )-\Tr(A_\text{diag}^\dag \rho_\text{eq} A_\text{diag})\ \\ &= \Tr(A^\dag \rho_\text{eq} A )-\Tr(A_\text{diag}^\dag \rho_\text{eq} A_\text{diag})\ .
\end{align}
\ees
Similarly we can show 
\begin{align}\label{sign1}
q_{A\overline{A}}(\omega)= \int dt \ e^{i\omega t}\ \Tr(\rho_\E A(t)\overline{A}^\dag)=p^\E_{A}(\omega) \times \text{Sign}(\omega)\ .
\end{align}
The latter equation implies that for odd integer $n$, 
\begin{align}\label{odd145}
\int d\omega\ \omega^n\  q_{A\overline{A}}(\omega)= \int d\omega\ \omega^n\  p^\E_{A}(\omega) \times \text{Sign}(\omega)=\int d\omega\ |\omega|^n\  p^\E_{A}(\omega) .
\end{align}
Finally, putting $B=\overline{A}$ in Eq.(\ref{formula1}) and using Eq.(\ref{odd145}) and Eq.(\ref{sign14}), together with the fact that $ \overline{A}_\text{diag}= \hat{A}(0) \text{Sign}(0)=0$,  we find that for odd integer $n$ in the interval $1 \le n\le l$,
\bes
\begin{align}
\int d\omega\ |\omega|^n\  p^\E_{A}(\omega)&= \int d\omega\ \omega^n\  q_{A\overline{A}}(\omega)  \\ &\le   2\pi  \sqrt{\Tr(\overline{A}^\dag \rho_\text{eq} \overline{A} )-\Tr(\overline{A}_\text{diag}^\dag \rho_\text{eq} \overline{A}_\text{diag})} \times  (4rk J_\text{max})^n \times n! \times 2^\frac{R_A}{rk}  \times \|A\|\\ &= 2\pi  \sqrt{\Tr(A^\dag \rho_\text{eq} A )-\Tr(A_\text{diag}^\dag \rho_\text{eq} A_\text{diag})} \times  (4rk J_\text{max})^n \times n! \times 2^\frac{R_A}{rk}  \times \|A\| \  .
\end{align}
\ees
This together with Eq.(\ref{evencase}) implies that for all integer $n$ in the interval $1\le n\le l$ it holds that
\begin{align}\label{odd45}
\int d\omega\ |\omega|^n\  p^\E_{A}(\omega)&\le  2\pi  \sqrt{\Tr(A^\dag \rho_\text{eq} A )-\Tr(A_\text{diag}^\dag \rho_\text{eq} A_\text{diag})} \times  (4rk J_\text{max})^n \times n! \times 2^\frac{R_A}{rk}  \times \|A\|\ .
\end{align}
This proves Eq.(\ref{main_app}).

Next, to prove Eq.(\ref{main200_app}) (or equivalently Eq.(9) in the main text) we multiply both sides of inequality (\ref{odd45}) in $\alpha^n/n!$, where $\alpha=(8rk J_\text{max})^{-1}  $,  and sum over $n$ from $1$ to $l$: 
\begin{align}
\int d\omega\ \sum_{n=1}^l \frac{\alpha^n}{n!} |\omega|^n\  p^\E_{A}(\omega)&\le      2\pi  \sqrt{\Tr(A^\dag \rho_\text{eq} A )-\Tr(A_\text{diag}^\dag \rho_\text{eq} A_\text{diag})} \times   2^\frac{R_A}{rk}  \times \|A\| \times\sum_{n=1}^l (4\alpha rk J_\text{max})^n\\ &\le      2\pi  \sqrt{\Tr(A^\dag \rho_\text{eq} A )-\Tr(A_\text{diag}^\dag \rho_\text{eq} A_\text{diag})} \times   2^\frac{R_A}{rk}  \times \|A\| \times\sum_{n=1}^l \frac{1}{2^n}
\\ &\le     2\pi  \sqrt{\Tr(A^\dag \rho_\text{eq} A )-\Tr(A_\text{diag}^\dag \rho_\text{eq} A_\text{diag})} \times   2^\frac{R_A}{rk}  \times \|A\|   .
\end{align}
Then, we use the facts that  both functions $p^\E_{A}(\omega)$ and $\sum_{n=1}^l \frac{(\alpha|\omega|)^n}{n!}$ are non-negative, and function $\sum_{n=1}^l \frac{(\alpha|\omega|)^n}{n!}$ is monotonically increasing with $|\omega|$. This implies
for any frequency  $\omega_0$, it holds that 
\bes\label{eee0}
\begin{align}
\Big( \sum_{n=1}^l \frac{\alpha^n |\omega_0|^n}{n!} \Big) P^\E_{A}(\omega_0)&=
 \Big( \sum_{n=1}^l \frac{\alpha^n |\omega_0|^n}{n!} \Big) \int_{|\omega|\ge |\omega_0|} d\omega\ p^\E_{A}(\omega) \\ &\le  \int d\omega \Big( \sum_{n=1}^l \frac{\alpha^n |\omega|^n}{n!} \Big) p^\E_{A}(\omega) 
\\ &\le      2\pi  \sqrt{\Tr(A^\dag \rho_\text{eq} A )-\Tr(A_\text{diag}^\dag \rho_\text{eq} A_\text{diag})} \times   2^\frac{R_A}{rk}  \times \|A\| \  .
\end{align}
\ees
Finally, we note that for any $\omega_0$,
\begin{align}\label{eee1}
P^\E_{A}(\omega_0)\le P^\E_{{A}}(0^+) \le  2\pi  \sqrt{\Tr(A^\dag \rho_\text{eq} A )-\Tr(A_\text{diag}^\dag \rho_\text{eq} A_\text{diag})} \times \|A\|  \  .
\end{align}
where the first inequality follows from the positivity of $p^\E_{A}(\omega)$, and the second inequality follows from Eq.(\ref{qrty}), together with the fact that $ \sqrt{\Tr(A^\dag \rho_\text{eq} A )-\Tr(A_\text{diag}^\dag \rho_\text{eq} A_\text{diag})}\le \sqrt{\Tr(A^\dag \rho_\text{eq} A )}\le\|A\|$.

Finally, adding both sides of Eq.(\ref{eee0}) and Eq.(\ref{eee1}) implies
\begin{align}
\Big( \sum_{n=0}^l \frac{\alpha^n |\omega_0|^n}{n!} \Big) P^\E_{A}(\omega_0)\le        2\pi  \sqrt{\Tr(A^\dag \rho_\text{eq} A )-\Tr(A_\text{diag}^\dag \rho_\text{eq} A_\text{diag})} \times   (2^\frac{R_A}{rk}+1)  \times \|A\| \  ,
\end{align}
Therefore,
\begin{align}
P^\E_{A}(\omega_0)&\le       2\pi  \sqrt{\Tr(A^\dag \rho_\text{eq} A )-\Tr(A_\text{diag}^\dag \rho_\text{eq} A_\text{diag})} \times \|A\| \times   \frac{(2^\frac{R_A}{rk}+1)}{\Big( \sum_{n=0}^l \frac{\alpha^n |\omega_0|^n}{n!} \Big)}\\ &=    2\pi  \sqrt{\Tr(A^\dag \rho_\text{eq} A )-\Tr(A_\text{diag}^\dag \rho_\text{eq} A_\text{diag})} \times \|A\| \times   \frac{2^\frac{R_A}{rk}+1}{\text{Exp}_l(\alpha \omega_0)}  \  .
\end{align}
This completes the proof.

\subsection{Proof of Eq.(\ref{counting})}\label{sec:lem}

Here we follow the argument of \cite{arad2014connecting}. Let $\text{ad}_{i}(\cdot)\equiv [h_i,\cdot]$. By expanding $\text{ad}_{H^R}=\sum_i \text{ad}_{i} $ we find
\beq\label{sum123}
\text{ad}^n_{H^R}(A)= \sum_{i_1,\cdots,i_n} \text{ad}_{i_n}\circ \text{ad}_{i_{n-1}} \cdots\circ \text{ad}_{i_{1}}(A) \ . 
\eeq
Using the assumption that $\|h_i\|\le J_\text{max}$, we find that for any operator $X$, $\|\text{ad}_{i}(X)\|\le 2 J_\text{max}\|X\|$. Therefore, each term in the above summation is bounded by
\beq
\left\|\text{ad}_{i_n}\circ \text{ad}_{i_{n-1}} \cdots\circ \text{ad}_{i_{1}}(A)\right\| \le \|A\|\times (2J_\text{max})^n \ .
\eeq
Furthermore, we can see that many terms in the summation in Eq.(\ref{sum123}) vanish. In particular, for any $r\le n$, $\text{ad}_{i_r}\circ \text{ad}_{i_{r-1}} \cdots\circ \text{ad}_{i_{1}}(A)$ is nonzero only if there are sites which are acted upon non-trivially by both  $h_{i_r}$ and $\text{ad}_{i_{r-1}} \cdots\circ \text{ad}_{i_{1}}(A)$. Based on this observation, in the following we find an upper bound on the number of nonzero terms in the summation in Eq.(\ref{sum123}).

First consider $\text{ad}_{i_r}\circ \text{ad}_{i_{r-1}} \cdots\circ \text{ad}_{i_{1}}(A)$ for $r=1$.  By assumption  $\text{ad}_{i_{1}}(A)$ is nonzero for at most $R_A$ different $i_1$. Next, we consider $r=2$, and count the number of different $i_2$ for which $\text{ad}_{i_2}\circ \text{ad}_{i_{r_1}}(A)$ is nonzero for a fixed $i_1$. Since $h_i$ are all $k$-local, it follows that  the support of $\text{ad}_{i_{r_1}}(A)$ is restricted to the support of $A$ and, at most, $k$ other different sites. By assumption there are at most $R_A$ different $h_{i_2}$ which acts non-trivially on the support of $A$. Furthermore, there are at most $r\times k$ different  $h_{i_2}$  which act non-trivially on the extra $k$ sites in the support of   $\text{ad}_{i_{r_1}}(A)$. So, we find that for any fixed $h_{i_1}$, $\text{ad}_{i_2}\circ \text{ad}_{i_{r_1}}(A)$  can be nonzero for at most  $R_A+k r$ different ${i_2}$. Repeating  this argument we find that the number of nonzero terms in the above expansion is bounded by
\begin{align}
N_n=R_A(R_A+kr)\cdots \big(R_A+(n-1)kr\big)\ 
\end{align}
Next, we find an upper bound on $N_n$. Let $Z=R_A/(kr)$ and $\lceil Z \rceil$ be the smallest integer larger than  or equal to $Z$. Then, following \cite{arad2014connecting}, we rewrite this as
\bes
\begin{align}
N_n&=(kr)^n Z (Z+1)\cdots \big(Z+(n-1)\big)\ \\ &\le (kr)^n \lceil Z \rceil (\lceil Z \rceil+1)\cdots \big(\lceil Z \rceil+(n-1)\big)\\ &=(kr)^n\times {\lceil Z \rceil+(n-1) \choose n}\times n!
\\ &\le(kr)^n\times 2^{\lceil Z \rceil+(n-1)}\times n!
\\ &\le(kr)^n\times 2^{Z+n}\times n!\ ,
\end{align}
\ees
where the fourth line follows from the binomial expansion. Therefore, we find
\begin{align}
N_n=R_A(R_A+kr)\cdots \big(R_A+(n-1)kr\big)\ &\le (2kr)^n \times n! \times 2^\frac{R_A}{rk}\ .
\end{align}
Finally, we note that the right-hand side of Eq.(\ref{sum123}) has at most $N_n$ nonzero terms, and each term is bounded by $(2J_\max)^n \|A\|$. Therefore,
\beq
\| \text{ad}^n_{H^R}(A)\| \le N_n \|A\|\times  (2J_\text{max})^n\le  (4  J_\text{max} kr)^n \times n! \times 2^\frac{R_A}{rk} .
\eeq

\newpage

\section{Proof of theorem 2 in the paper}
Theorem 1 implies that if the error detection condition holds, then 
\begin{align}\label{lem-app}
\underset{\rho_S\in \mathcal{S}(\mathcal{C})}{\min} F^2(\rho_\S, U_\LS^\dag(t)\rho_\S(t) U_\LS(t))\nonumber=  1-\lambda^2\Big\| \sum_{E_n\ge E_\text{gap}} b_{n}(t)\ \Pi_\mathcal{C}  S \Pi_{E_n} S\Pi_\mathcal{C}\Big\| +\mathcal{O}(\lambda^3)\ ,
\end{align}
where $b_{n}(t)= \frac{1}{2\pi}\int d\omega\ p_B^\E(\omega)\left[\frac{\sin (\omega+E_n)t/2}{(\omega+E_n)/2}\right]^2$. Using the facts that  $b_n(t)$ are positive functions, and $\Pi_\mathcal{C}  S \Pi_{E_n} S\Pi_\mathcal{C}$ are positive operators we find
\bes
\begin{align}
\Big\| \sum_{E_n\ge E_\text{gap}} b_{n}(t)\ \Pi_\mathcal{C}  S \Pi_{E_n} S\Pi_\mathcal{C}\Big\| &=\underset{\psi}{\max}  \sum_{E_n\ge E_\text{gap}} b_{n}(t) \ \langle\psi|\Pi_\mathcal{C}  S \Pi_{E_n} S\Pi_\mathcal{C}|\psi\rangle\\ &\le\underset{E_n\ge E_\text{gap}}{\max} b_n(t)\times \underset{\psi}{\max} \sum_{E_n\ge E_\text{gap}}  \ \langle\psi|\Pi_\mathcal{C}  S \Pi_{E_n} S\Pi_\mathcal{C}|\psi\rangle\\  &\le \underset{E_n\ge E_\text{gap}}{\max} b_n(t)\times \left\| \Pi_\mathcal{C}  S^2 \Pi_\mathcal{C}\right\|\ ,
\end{align}
\ees
where to get the second  line we have used the positivity of  $b_n(t)$ and $\Pi_\mathcal{C}  S \Pi_{E_n} S\Pi_\mathcal{C}$. Putting this in Eq.(\ref{lem-app}) we find 
\begin{align}\label{ee1}
1-\underset{\rho_S\in \mathcal{S}(\mathcal{C})}{\min} F^2(\rho_\S, U_\LS^\dag(t)\rho_\S(t) U_\LS(t))\le\lambda^2  \left\| \Pi_\mathcal{C}  S^2 \Pi_\mathcal{C}\right\| \times \underset{E_n\ge E_\text{gap}}{\max} b_n(t)+\mathcal{O}(\lambda^3)\ .
\end{align}
Next, we note that for $E_n\ge E_\text{gap}$ it holds that
\bes\label{ee2}
\begin{align}
b_n(t) &= \frac{1}{2\pi}\int d\omega\ p_B^\E(\omega)\ \left[\frac{\sin (\omega+E_n)t/2}{(\omega+E_n)/2}\right]^2\\ &= \frac{1}{2\pi} \int_{-\infty}^{-\frac{E_\text{gap}}{2}} d\omega\ p_B^\E(\omega)\ \left[\frac{\sin (\omega+E_n)t/2}{(\omega+E_n)/2}\right]^2 + \frac{1}{2\pi}\int_{-\frac{E_\text{gap}}{2}}^{\infty} d\omega\ p_B^\E(\omega)\ \left[\frac{\sin (\omega+E_n)t/2}{(\omega+E_n)/2}\right]^2 \\ &\le    \frac{t^2}{2\pi}\int_{-\infty}^{-\frac{E_\text{gap}}{2}} d\omega\ p_B^\E(\omega) + \frac{16}{E^2_\text{gap}\times 2\pi} \int_{-\frac{E_\text{gap}}{2}}^{\infty} d\omega\ p_B^\E(\omega) \\ &\le    \frac{t^2}{2\pi} P_B^\E(\frac{E_\text{gap}}{2}) + \frac{16}{2\pi E^2_\text{gap}} \int d\omega\ p_B^\E(\omega) \\ &\le  \frac{t^2}{2\pi} P_B^\E(\frac{E_\text{gap}}{2}) +\frac{16 \|B\|^2}{E^2_\text{gap}} \ ,
\end{align}
\ees
where to get the second line we have decomposed the integral as the sum of two integrals over the intervals $(-\infty,-E_\text{gap}/2)$ and $[-E_\text{gap}/2,\infty)$, to get the third line we have used the fact that $(\frac{\sin (\omega+E_n)t/2}{(\omega+E_n)/2})^2$ is bounded by $t^2$ in the first interval and by $16/E^2_\text{gap}$ in the second interval (note that $E_n\ge E_\text{gap}$), to get the fourth line we have used the  positivity of $p_B^\E(\omega)$ and the definition 
\beq
P_B^\E(\omega)=\int_{-\infty}^{-|\omega|} d\nu\ p_B^\E(\nu)+ \int_{|\omega|}^{\infty} d\nu\ p_B^\E(\nu)\ ,
\eeq
and to get the last line we have used $\int d\omega\ p^\E_B(\omega)=2\pi \Tr(\rho_\B B^2)\le  2\pi \|B\|^2$.

Finally, using theorem \ref{Thm:app} in the supplementary material, or Eq.(9) in the  paper, we have
\bes\label{ee3}
\begin{align}
P^\E_B(\frac{E_\text{gap}}{2})&\le \|B\| \sqrt{2\pi P^\E_B(0^+)} \times \frac{2^\frac{R_\I}{rk}+1}{\text{Exp}_l\big(\frac{E_\text{gap}}{16 J_\text{max} r k} \big) } \\ &\le \|B\|^2 (2\pi) \times \frac{2^\frac{R_\I}{rk}+1}{\text{Exp}_l\big(\frac{E_\text{gap}}{16 J_\text{max} r k} \big) }\ ,
\end{align}
\ees
where we have used the fact the total AC power satisfies 
\beq
P^\E_B(0^+)\le \int d\omega\ p^\E_B(\omega)=2\pi \Tr(\rho_\B B^2)\le  2\pi \|B\|^2\ .
\eeq
Putting Eq.(\ref{ee3}) into Eq.(\ref{ee2}) we find 
\begin{align}
b_n(t) &\le  \|B\|^2 \Big(  t^2  \frac{2^\frac{R_\I}{rk}+1}{\text{Exp}_l\big(\frac{E_\text{gap}}{16 J_\text{max} r k} \big) } + \frac{16 }{E^2_\text{gap}}\Big)\ .
\end{align}
This together with Eq.(\ref{ee1}) implies that
\bes
\begin{align}
1-\underset{\rho_S\in \mathcal{S}(\mathcal{C})}{\min} F^2(\rho_\S, U_\LS^\dag(t)\rho_\S(t) U_\LS(t))&\le\lambda^2  \left\| \Pi_\mathcal{C}  S^2 \Pi_\mathcal{C}\right\| \times \underset{E_n\ge E_\text{gap}}{\max} b_n(t)+\mathcal{O}(\lambda^3)\\ &\le \lambda^2  \left\| \Pi_\mathcal{C}  S^2 \Pi_\mathcal{C}\right\| \times \|B\|^2 \Big(  t^2  \frac{2^\frac{R_\I}{rk}+1}{\text{Exp}_l\big(\frac{E_\text{gap}}{16 J_\text{max} r k} \big) } + \frac{16 }{E^2_\text{gap}}\Big) +\mathcal{O}(\lambda^3)\\  &= \lambda^2  \left\| \Pi_\mathcal{C}  H_\I^2 \Pi_\mathcal{C}\right\| \Big(  t^2  \frac{2^\frac{R_\I}{rk}+1}{\text{Exp}_l\big(\frac{E_\text{gap}}{16 J_\text{max} r k} \big) } + \frac{16 }{E^2_\text{gap}}\Big) +\mathcal{O}(\lambda^3) ,
\end{align}
\ees
where we have used $ \left\| \Pi_\mathcal{C}  H_\I^2 \Pi_\mathcal{C}\right\|= \left\| \Pi_\mathcal{C}  S^2 \Pi_\mathcal{C}\otimes B^2\right\|=\| \Pi_\mathcal{C}  S^2 \Pi_\mathcal{C}\|\times \|B^2\|=\| \Pi_\mathcal{C}  S^2 \Pi_\mathcal{C}\|\times \|B\|^2$. This completes the proof of theorem 2.

\newpage

\bibliography{ref}

\begin{thebibliography}{30}%
\makeatletter
\providecommand \@ifxundefined [1]{%
 \@ifx{#1\undefined}
}%
\providecommand \@ifnum [1]{%
 \ifnum #1\expandafter \@firstoftwo
 \else \expandafter \@secondoftwo
 \fi
}%
\providecommand \@ifx [1]{%
 \ifx #1\expandafter \@firstoftwo
 \else \expandafter \@secondoftwo
 \fi
}%
\providecommand \natexlab [1]{#1}%
\providecommand \enquote  [1]{``#1''}%
\providecommand \bibnamefont  [1]{#1}%
\providecommand \bibfnamefont [1]{#1}%
\providecommand \citenamefont [1]{#1}%
\providecommand \href@noop [0]{\@secondoftwo}%
\providecommand \href [0]{\begingroup \@sanitize@url \@href}%
\providecommand \@href[1]{\@@startlink{#1}\@@href}%
\providecommand \@@href[1]{\endgroup#1\@@endlink}%
\providecommand \@sanitize@url [0]{\catcode `\\12\catcode `\$12\catcode
  `\&12\catcode `\#12\catcode `\^12\catcode `\_12\catcode `\%12\relax}%
\providecommand \@@startlink[1]{}%
\providecommand \@@endlink[0]{}%
\providecommand \url  [0]{\begingroup\@sanitize@url \@url }%
\providecommand \@url [1]{\endgroup\@href {#1}{\urlprefix }}%
\providecommand \urlprefix  [0]{URL }%
\providecommand \Eprint [0]{\href }%
\providecommand \doibase [0]{http://dx.doi.org/}%
\providecommand \selectlanguage [0]{\@gobble}%
\providecommand \bibinfo  [0]{\@secondoftwo}%
\providecommand \bibfield  [0]{\@secondoftwo}%
\providecommand \translation [1]{[#1]}%
\providecommand \BibitemOpen [0]{}%
\providecommand \bibitemStop [0]{}%
\providecommand \bibitemNoStop [0]{.\EOS\space}%
\providecommand \EOS [0]{\spacefactor3000\relax}%
\providecommand \BibitemShut  [1]{\csname bibitem#1\endcsname}%
\let\auto@bib@innerbib\@empty
\bibitem [{\citenamefont {Shor}(1995)}]{shor1995scheme}%
  \BibitemOpen
  \bibfield  {author} {\bibinfo {author} {\bibfnamefont {P.~W.}\ \bibnamefont
  {Shor}},\ }\href@noop {} {\bibfield  {journal} {\bibinfo  {journal} {Physical
  review A}\ }\textbf {\bibinfo {volume} {52}},\ \bibinfo {pages} {R2493}
  (\bibinfo {year} {1995})}\BibitemShut {NoStop}%
\bibitem [{\citenamefont {{D. Gottesman}}(1997)}]{Gottesman:97b}%
  \BibitemOpen
  \bibfield  {author} {\bibinfo {author} {\bibnamefont {{D. Gottesman}}},\
  }\emph {\bibinfo {title} {{Stabilizer Codes and Quantum Error Correction}}},\
  \href@noop {} {Ph.D. thesis},\ \bibinfo  {school} {California Institute of
  Technology}, \bibinfo {address} {Pasadena, CA} (\bibinfo {year} {1997}),\
  \Eprint {http://arxiv.org/abs/quant-ph/9705052} {quant-ph/9705052}
  \BibitemShut {NoStop}%
\bibitem [{\citenamefont {Zanardi}\ and\ \citenamefont
  {Rasetti}(1997)}]{Zanardi:97c}%
  \BibitemOpen
  \bibfield  {author} {\bibinfo {author} {\bibfnamefont {P.}~\bibnamefont
  {Zanardi}}\ and\ \bibinfo {author} {\bibfnamefont {M.}~\bibnamefont
  {Rasetti}},\ }\href@noop {} {\bibfield  {journal} {\bibinfo  {journal} {Phys.
  Rev. Lett.}\ }\textbf {\bibinfo {volume} {79}},\ \bibinfo {pages} {3306}
  (\bibinfo {year} {1997})}\BibitemShut {NoStop}%
\bibitem [{\citenamefont {{D. A. Lidar, I. L. Chuang, and K. B.
  Whaley}}(1998)}]{LidarDFS}%
  \BibitemOpen
  \bibfield  {author} {\bibinfo {author} {\bibnamefont {{D. A. Lidar, I. L.
  Chuang, and K. B. Whaley}}},\ }\href@noop {} {\bibfield  {journal} {\bibinfo
  {journal} {Phys. Rev. Lett.}\ }\textbf {\bibinfo {volume} {81}},\ \bibinfo
  {pages} {2594} (\bibinfo {year} {1998})}\BibitemShut {NoStop}%
\bibitem [{\citenamefont {Viola}\ \emph {et~al.}(1999)\citenamefont {Viola},
  \citenamefont {Knill},\ and\ \citenamefont {Lloyd}}]{Viola:99}%
  \BibitemOpen
  \bibfield  {author} {\bibinfo {author} {\bibfnamefont {L.}~\bibnamefont
  {Viola}}, \bibinfo {author} {\bibfnamefont {E.}~\bibnamefont {Knill}}, \ and\
  \bibinfo {author} {\bibfnamefont {S.}~\bibnamefont {Lloyd}},\ }\href@noop {}
  {\bibfield  {journal} {\bibinfo  {journal} {Phys. Rev. Lett.}\ }\textbf
  {\bibinfo {volume} {82}},\ \bibinfo {pages} {2417} (\bibinfo {year}
  {1999})}\BibitemShut {NoStop}%
\bibitem [{\citenamefont {Viola}\ and\ \citenamefont {Lloyd}(1998)}]{Viola:98}%
  \BibitemOpen
  \bibfield  {author} {\bibinfo {author} {\bibfnamefont {L.}~\bibnamefont
  {Viola}}\ and\ \bibinfo {author} {\bibfnamefont {S.}~\bibnamefont {Lloyd}},\
  }\href@noop {} {\bibfield  {journal} {\bibinfo  {journal} {Phys. Rev. A}\
  }\textbf {\bibinfo {volume} {58}},\ \bibinfo {pages} {2733} (\bibinfo {year}
  {1998})}\BibitemShut {NoStop}%
\bibitem [{\citenamefont {Lidar}(2008)}]{PhysRevLett.100.160506}%
  \BibitemOpen
  \bibfield  {author} {\bibinfo {author} {\bibfnamefont {D.~A.}\ \bibnamefont
  {Lidar}},\ }\href {http://link.aps.org/doi/10.1103/PhysRevLett.100.160506}
  {\bibfield  {journal} {\bibinfo  {journal} {{Phys.~Rev.~Lett.}}\ }\textbf
  {\bibinfo {volume} {100}},\ \bibinfo {pages} {160506} (\bibinfo {year}
  {2008})}\BibitemShut {NoStop}%
\bibitem [{\citenamefont {Kitaev}(2003)}]{kitaev2003fault}%
  \BibitemOpen
  \bibfield  {author} {\bibinfo {author} {\bibfnamefont {A.~Y.}\ \bibnamefont
  {Kitaev}},\ }\href@noop {} {\bibfield  {journal} {\bibinfo  {journal} {Annals
  of Physics}\ }\textbf {\bibinfo {volume} {303}},\ \bibinfo {pages} {2}
  (\bibinfo {year} {2003})}\BibitemShut {NoStop}%
\bibitem [{\citenamefont {Zanardi}\ and\ \citenamefont {Rasetti}(1999)}]{HQC}%
  \BibitemOpen
  \bibfield  {author} {\bibinfo {author} {\bibfnamefont {P.}~\bibnamefont
  {Zanardi}}\ and\ \bibinfo {author} {\bibfnamefont {M.}~\bibnamefont
  {Rasetti}},\ }\href {\doibase
  http://dx.doi.org/10.1016/S0375-9601(99)00803-8} {\bibfield  {journal}
  {\bibinfo  {journal} {Physics Letters A}\ }\textbf {\bibinfo {volume}
  {264}},\ \bibinfo {pages} {94} (\bibinfo {year} {1999})}\BibitemShut
  {NoStop}%
\bibitem [{\citenamefont {Jordan}\ \emph {et~al.}(2006)\citenamefont {Jordan},
  \citenamefont {Farhi},\ and\ \citenamefont {Shor}}]{PhysRevA.74.052322}%
  \BibitemOpen
  \bibfield  {author} {\bibinfo {author} {\bibfnamefont {S.~P.}\ \bibnamefont
  {Jordan}}, \bibinfo {author} {\bibfnamefont {E.}~\bibnamefont {Farhi}}, \
  and\ \bibinfo {author} {\bibfnamefont {P.~W.}\ \bibnamefont {Shor}},\ }\href
  {\doibase 10.1103/PhysRevA.74.052322} {\bibfield  {journal} {\bibinfo
  {journal} {Phys. Rev. A}\ }\textbf {\bibinfo {volume} {74}},\ \bibinfo
  {pages} {052322} (\bibinfo {year} {2006})}\BibitemShut {NoStop}%
\bibitem [{\citenamefont {Farhi}\ \emph {et~al.}(2000)\citenamefont {Farhi},
  \citenamefont {Goldstone}, \citenamefont {Gutmann},\ and\ \citenamefont
  {Sipser}}]{FarhiAQC:00}%
  \BibitemOpen
  \bibfield  {author} {\bibinfo {author} {\bibfnamefont {E.}~\bibnamefont
  {Farhi}}, \bibinfo {author} {\bibfnamefont {J.}~\bibnamefont {Goldstone}},
  \bibinfo {author} {\bibfnamefont {S.}~\bibnamefont {Gutmann}}, \ and\
  \bibinfo {author} {\bibfnamefont {M.}~\bibnamefont {Sipser}},\ }\href
  {http://arxiv.org/abs/quant-ph/0001106} {\bibfield  {journal} {\bibinfo
  {journal} {ArXiv}\ } (\bibinfo {year} {2000})},\ \Eprint
  {http://arxiv.org/abs/quant-ph/0001106 (2000)} {quant-ph/0001106 (2000)}
  \BibitemShut {NoStop}%
\bibitem [{\citenamefont {Lidar}\ and\ \citenamefont
  {Brun}(2013)}]{Lidar-Brun:book}%
  \BibitemOpen
  \bibinfo {editor} {\bibfnamefont {D.}~\bibnamefont {Lidar}}\ and\ \bibinfo
  {editor} {\bibfnamefont {T.}~\bibnamefont {Brun}},\ eds.,\ \href
  {http://www.cambridge.org/9780521897877} {\emph {\bibinfo {title} {Quantum
  Error Correction}}}\ (\bibinfo  {publisher} {Cambridge University Press},\
  \bibinfo {address} {{Cambride, UK}},\ \bibinfo {year} {2013})\BibitemShut
  {NoStop}%
\bibitem [{\citenamefont {Knill}\ \emph {et~al.}(2000)\citenamefont {Knill},
  \citenamefont {Laflamme},\ and\ \citenamefont {Viola}}]{knill2000theory}%
  \BibitemOpen
  \bibfield  {author} {\bibinfo {author} {\bibfnamefont {E.}~\bibnamefont
  {Knill}}, \bibinfo {author} {\bibfnamefont {R.}~\bibnamefont {Laflamme}}, \
  and\ \bibinfo {author} {\bibfnamefont {L.}~\bibnamefont {Viola}},\
  }\href@noop {} {\bibfield  {journal} {\bibinfo  {journal} {Physical Review
  Letters}\ }\textbf {\bibinfo {volume} {84}},\ \bibinfo {pages} {2525}
  (\bibinfo {year} {2000})}\BibitemShut {NoStop}%
\bibitem [{\citenamefont {Bookatz}\ \emph {et~al.}(2015)\citenamefont
  {Bookatz}, \citenamefont {Farhi},\ and\ \citenamefont
  {Zhou}}]{bookatz2015error}%
  \BibitemOpen
  \bibfield  {author} {\bibinfo {author} {\bibfnamefont {A.~D.}\ \bibnamefont
  {Bookatz}}, \bibinfo {author} {\bibfnamefont {E.}~\bibnamefont {Farhi}}, \
  and\ \bibinfo {author} {\bibfnamefont {L.}~\bibnamefont {Zhou}},\ }\href@noop
  {} {\bibfield  {journal} {\bibinfo  {journal} {Physical Review A}\ }\textbf
  {\bibinfo {volume} {92}},\ \bibinfo {pages} {022317} (\bibinfo {year}
  {2015})}\BibitemShut {NoStop}%
\bibitem [{\citenamefont {Marvian}\ and\ \citenamefont
  {Lidar}(2015)}]{marvian2015quantum}%
  \BibitemOpen
  \bibfield  {author} {\bibinfo {author} {\bibfnamefont {I.}~\bibnamefont
  {Marvian}}\ and\ \bibinfo {author} {\bibfnamefont {D.~A.}\ \bibnamefont
  {Lidar}},\ }\href@noop {} {\bibfield  {journal} {\bibinfo  {journal}
  {Physical review letters}\ }\textbf {\bibinfo {volume} {115}},\ \bibinfo
  {pages} {210402} (\bibinfo {year} {2015})}\BibitemShut {NoStop}%
\bibitem [{\citenamefont {Marvian}\ and\ \citenamefont
  {Lidar}(2014)}]{marvian2014quantum}%
  \BibitemOpen
  \bibfield  {author} {\bibinfo {author} {\bibfnamefont {I.}~\bibnamefont
  {Marvian}}\ and\ \bibinfo {author} {\bibfnamefont {D.~A.}\ \bibnamefont
  {Lidar}},\ }\href@noop {} {\bibfield  {journal} {\bibinfo  {journal}
  {Physical review letters}\ }\textbf {\bibinfo {volume} {113}},\ \bibinfo
  {pages} {260504} (\bibinfo {year} {2014})}\BibitemShut {NoStop}%
\bibitem [{\citenamefont {{N.V. Prokof'ev and P.C.E.
  Stamp}}(2000)}]{Prokofev:00}%
  \BibitemOpen
  \bibfield  {author} {\bibinfo {author} {\bibnamefont {{N.V. Prokof'ev and
  P.C.E. Stamp}}},\ }\href@noop {} {\bibfield  {journal} {\bibinfo  {journal}
  {{Rep. Prog. Phys.}}\ }\textbf {\bibinfo {volume} {63}},\ \bibinfo {pages}
  {669} (\bibinfo {year} {2000})}\BibitemShut {NoStop}%
\bibitem [{\citenamefont {{R. Alicki, M. Horodecki, P. Horodecki, and R.
  Horodecki}}(2002)}]{Alicki:02}%
  \BibitemOpen
  \bibfield  {author} {\bibinfo {author} {\bibnamefont {{R. Alicki, M.
  Horodecki, P. Horodecki, and R. Horodecki}}},\ }\href@noop {} {\bibfield
  {journal} {\bibinfo  {journal} {Phys. Rev. A}\ }\textbf {\bibinfo {volume}
  {65}},\ \bibinfo {pages} {062101} (\bibinfo {year} {2002})}\BibitemShut
  {NoStop}%
\bibitem [{\citenamefont {Alicki}(1989)}]{PhysRevA.40.4077}%
  \BibitemOpen
  \bibfield  {author} {\bibinfo {author} {\bibfnamefont {R.}~\bibnamefont
  {Alicki}},\ }\href {\doibase 10.1103/PhysRevA.40.4077} {\bibfield  {journal}
  {\bibinfo  {journal} {Phys. Rev. A}\ }\textbf {\bibinfo {volume} {40}},\
  \bibinfo {pages} {4077} (\bibinfo {year} {1989})}\BibitemShut {NoStop}%
\bibitem [{\citenamefont {Majenz}\ \emph {et~al.}(2013)\citenamefont {Majenz},
  \citenamefont {Albash}, \citenamefont {Breuer},\ and\ \citenamefont
  {Lidar}}]{Majenz:2013qw}%
  \BibitemOpen
  \bibfield  {author} {\bibinfo {author} {\bibfnamefont {C.}~\bibnamefont
  {Majenz}}, \bibinfo {author} {\bibfnamefont {T.}~\bibnamefont {Albash}},
  \bibinfo {author} {\bibfnamefont {H.-P.}\ \bibnamefont {Breuer}}, \ and\
  \bibinfo {author} {\bibfnamefont {D.~A.}\ \bibnamefont {Lidar}},\ }\href
  {http://link.aps.org/doi/10.1103/PhysRevA.88.012103} {\bibfield  {journal}
  {\bibinfo  {journal} {Physical Review A}\ }\textbf {\bibinfo {volume} {88}},\
  \bibinfo {pages} {012103} (\bibinfo {year} {2013})}\BibitemShut {NoStop}%
\bibitem [{\citenamefont {Nielsen}\ and\ \citenamefont
  {Chuang}(2000)}]{nielsen2000quantum}%
  \BibitemOpen
  \bibfield  {author} {\bibinfo {author} {\bibfnamefont {M.}~\bibnamefont
  {Nielsen}}\ and\ \bibinfo {author} {\bibfnamefont {I.}~\bibnamefont
  {Chuang}},\ }\href@noop {} {\emph {\bibinfo {title} {Quantum Computation and
  Quantum Information}}},\ Cambridge Series on Information and the Natural
  Sciences\ (\bibinfo  {publisher} {Cambridge University Press},\ \bibinfo
  {year} {2000})\BibitemShut {NoStop}%
\bibitem [{\citenamefont {Uhlmann}(1976)}]{Uhlmann}%
  \BibitemOpen
  \bibfield  {author} {\bibinfo {author} {\bibfnamefont {A.}~\bibnamefont
  {Uhlmann}},\ }\href@noop {} {\bibfield  {journal} {\bibinfo  {journal}
  {Reports on Mathematical Physics}\ }\textbf {\bibinfo {volume} {9}},\
  \bibinfo {pages} {273} (\bibinfo {year} {1976})}\BibitemShut {NoStop}%
\bibitem [{\citenamefont {Jozsa}(1994)}]{Fidelity_Jozsa}%
  \BibitemOpen
  \bibfield  {author} {\bibinfo {author} {\bibfnamefont {R.}~\bibnamefont
  {Jozsa}},\ }\href@noop {} {\bibfield  {journal} {\bibinfo  {journal} {Journal
  of Modern Optics}\ }\textbf {\bibinfo {volume} {41}},\ \bibinfo {pages}
  {2315} (\bibinfo {year} {1994})}\BibitemShut {NoStop}%
\bibitem [{\citenamefont {{E. H. Lieb and D. W.
  Robinson}}(1972)}]{LiebRobinson}%
  \BibitemOpen
  \bibfield  {author} {\bibinfo {author} {\bibnamefont {{E. H. Lieb and D. W.
  Robinson}}},\ }\href@noop {} {\bibfield  {journal} {\bibinfo  {journal}
  {Comm. Math. Phys.}\ }\textbf {\bibinfo {volume} {28}},\ \bibinfo {pages}
  {251} (\bibinfo {year} {1972})}\BibitemShut {NoStop}%
\bibitem [{\citenamefont {Hastings}(2010)}]{hastings2010locality}%
  \BibitemOpen
  \bibfield  {author} {\bibinfo {author} {\bibfnamefont {M.~B.}\ \bibnamefont
  {Hastings}},\ }\href@noop {} {\bibfield  {journal} {\bibinfo  {journal}
  {arXiv preprint arXiv:1008.5137}\ } (\bibinfo {year} {2010})}\BibitemShut
  {NoStop}%
\bibitem [{\citenamefont {Hastings}(2007)}]{Hastings_area}%
  \BibitemOpen
  \bibfield  {author} {\bibinfo {author} {\bibfnamefont {M.~B.}\ \bibnamefont
  {Hastings}},\ }\href@noop {} {\bibfield  {journal} {\bibinfo  {journal}
  {Journal of Statistical Mechanics: Theory and Experiment}\ }\textbf {\bibinfo
  {volume} {2007}},\ \bibinfo {pages} {P08024} (\bibinfo {year}
  {2007})}\BibitemShut {NoStop}%
\bibitem [{\citenamefont {Bravyi}\ \emph {et~al.}(2006)\citenamefont {Bravyi},
  \citenamefont {Hastings},\ and\ \citenamefont {Verstraete}}]{bravyi2006lieb}%
  \BibitemOpen
  \bibfield  {author} {\bibinfo {author} {\bibfnamefont {S.}~\bibnamefont
  {Bravyi}}, \bibinfo {author} {\bibfnamefont {M.}~\bibnamefont {Hastings}}, \
  and\ \bibinfo {author} {\bibfnamefont {F.}~\bibnamefont {Verstraete}},\
  }\href@noop {} {\bibfield  {journal} {\bibinfo  {journal} {Physical review
  letters}\ }\textbf {\bibinfo {volume} {97}},\ \bibinfo {pages} {050401}
  (\bibinfo {year} {2006})}\BibitemShut {NoStop}%
\bibitem [{\citenamefont {Arad}\ \emph {et~al.}(2014)\citenamefont {Arad},
  \citenamefont {Kuwahara},\ and\ \citenamefont {Landau}}]{arad2014connecting}%
  \BibitemOpen
  \bibfield  {author} {\bibinfo {author} {\bibfnamefont {I.}~\bibnamefont
  {Arad}}, \bibinfo {author} {\bibfnamefont {T.}~\bibnamefont {Kuwahara}}, \
  and\ \bibinfo {author} {\bibfnamefont {Z.}~\bibnamefont {Landau}},\
  }\href@noop {} {\bibfield  {journal} {\bibinfo  {journal} {arXiv preprint
  arXiv:1406.3898}\ } (\bibinfo {year} {2014})}\BibitemShut {NoStop}%
\bibitem [{\citenamefont {Marvian}(2016)}]{prep}%
  \BibitemOpen
  \bibfield  {author} {\bibinfo {author} {\bibfnamefont {I.}~\bibnamefont
  {Marvian}},\ }\href@noop {} {\emph {\bibinfo {title} {Under preparation}}}\
  (\bibinfo {year} {2016})\BibitemShut {NoStop}%
\bibitem [{\citenamefont {Young}\ \emph {et~al.}(2013)\citenamefont {Young},
  \citenamefont {Sarovar},\ and\ \citenamefont {Blume-Kohout}}]{Young:13}%
  \BibitemOpen
  \bibfield  {author} {\bibinfo {author} {\bibfnamefont {K.~C.}\ \bibnamefont
  {Young}}, \bibinfo {author} {\bibfnamefont {M.}~\bibnamefont {Sarovar}}, \
  and\ \bibinfo {author} {\bibfnamefont {R.}~\bibnamefont {Blume-Kohout}},\
  }\href {http://link.aps.org/doi/10.1103/PhysRevX.3.041013} {\bibfield
  {journal} {\bibinfo  {journal} {Physical Review X}\ }\textbf {\bibinfo
  {volume} {3}},\ \bibinfo {pages} {041013} (\bibinfo {year}
  {2013})}\BibitemShut {NoStop}%
\end{thebibliography}%

\end{document}